\documentclass{article}

\usepackage{arxiv}

\usepackage[utf8]{inputenc} 
\usepackage[T1]{fontenc}    
\usepackage{hyperref}       
\usepackage{url}            
\usepackage{booktabs}       
\usepackage{amsfonts}       
\usepackage{nicefrac}       
\usepackage{microtype}      
\usepackage{lipsum}
\usepackage{amsmath}
\DeclareMathOperator*{\argminB}{argmin}
\usepackage{mathtools}
\DeclarePairedDelimiter\ceil{\lceil}{\rceil}
\DeclarePairedDelimiter\floor{\lfloor}{\rfloor}
\usepackage{multirow}
\usepackage[round]{natbib}
\usepackage{placeins}

\title{Higher Criticism Tuned Regression For Weak And Sparse Signals}

\author{
  Tao Jiang\\
  North Carolina State University\\
  \texttt{tjiang8@ncsu.edu} \\
  \And
  Stephanie J. London \\
  National Institute of Environmental Health Sciences\\
  \And
  Mi Kyeong Lee \\
  National Institute of Environmental Health Sciences\\
  \And
  Josyf C. Mychaleckyj \\
  Center for Public Health Genomics, University of Virginia \\
  \And
  Alison A. Motsinger-Reif\thanks{To whom correspondence should be addressed.}\\
  National Institute of Environmental Health Sciences\\
  \texttt{motsingerreifaa@nih.gov} \\
}

\begin{document}
\maketitle

\begin{abstract}
Here we propose a novel searching scheme for a tuning parameter in high-dimensional penalized regression methods to address variable selection and modeling when sample sizes are limited compared to the data dimensions. Our method is motivated by high-throughput biological data such as genome-wide association studies (GWAS) and epigenome-wide association studies (EWAS). We propose a new estimate of the regularization parameter $\lambda$ in penalized regression methods based on an estimated lower bound of the proportion of false null hypotheses with confidence $(1-\alpha)$. The bound is estimated by applying the empirical null distribution of the higher criticism statistic, a second-level significance test constructed by dependent \textit{p}-values using a multi-split regression and aggregation method. A tuning parameter estimate in penalized regression, $\lambda$, corresponds with the lower bound of the proportion of false null hypotheses. Different penalized regression methods with varied signal sparsity and strength are compared in the multi-split method setting. We demonstrate the performance of our method using both simulation experiments and the applications of real data on (1) lipid-trait genetics from the Action to Control Cardiovascular Risk in Diabetes (ACCORD) clinical trial and (2) epigenetic analysis evaluating smoking's influence in differential methylation in the Agricultural Lung Health Study. The proposed algorithm is included in the HCTR package, available at \url{https://cran.r-project.org/web/packages/HCTR/index.html}.
\end{abstract}


\section{Introduction}
High-throughput technologies in genetics and genomics present new challenges in high-dimensional data analysis. Genome-wide association studies (GWAS) have become a common tool to identify genetic loci associated with common complex diseases or disease-relevant phenotypes. It is understood that the etiology of many of the studied traits involves a large number of loci with small effects \citep{complex2003nature}. Most current methods, such as the mixed linear model approach \citep{yu2006unified,zhang2010mixed} and genome-wide efficient mixed-model association \citep{zhou2012genome}, require Bonferroni correction for multiple hypotheses after association analysis on a locus-by-locus basis (with millions of loci in each study). Assume $\alpha$ is the desired overall significance level and $p$ is the number of hypotheses. Bonferroni correction tests each individual hypothesis at a significance level of $\alpha/p$. However, in high-dimensional data, as $p \to \infty$ and $\alpha/p \to 0$, the Bonferroni correction is overly conservative \citep{perneger1998s} so that the power of GWAS is too low (even if correction is only for the effective number of loci after correcting for correlations). Therefore, important loci underlying complex traits are likely to be missed by traditional Bonferroni correction.

Multi-locus models of penalized regression, such as penalized logistic regression \citep{ayers2010snp} and elastic net \citep{cho2010joint}, have been proposed as alternative strategies to single-locus GWAS methods because of their natural properties without Bonferroni correction. Since these methods are shrinkage approaches, tuning of the regularization parameter $(\lambda)$ is extremely important \citep{tibshirani1996regression}. The accuracy of variable/predictor selection is highly related to the choice of the regularization parameter \citep{olson2017data}. The most widely accepted method for tuning the regularization parameter is described in \cite{friedman2010regularization}. This algorithm computes solutions for a decreasing sequence of values for $\lambda$. In this decreasing sequence, $\lambda_{\text{max}}$ is the smallest value that can shrink the entire coefficient estimations to zero, and $\lambda_{\text{min}}=c\lambda_{\text{max}}$, where $c$ is a small positive constant. The final choice of $\lambda$ in this sequence is data-driven and optimizes the $\lambda$ value that minimizes the value of the loss function. However, the tuning range and final decision on the regularization parameter lacks theoretical support from domain knowledge, such as genetics in GWAS.

Another limit of penalized regression is the requirement for the signal strength. Here, we define $S_0=\{ i; \beta_i^0 \neq 0, i=1,...,p \}$ as the set of true non-zero variables, where $\beta_i^0$ is the $i$-th variable in the true model, and the corresponding $s_0=| S_0 |$. \cite{van2011adaptive} proposed a beta-min condition to guarantee that all non-zero coefficients are sufficiently large
\begin{equation}
    \min_{i \in S_0} |\beta_i^0| \geq \beta_{min},
\end{equation}
for a positive constant $\beta_{min}$. One of such $\beta_{min}$s is $\phi^{-2} \sqrt{s_0 \log{(p)}/n}$, where $\phi^{-2}$ denotes a restricted eigenvalue of the design matrix. Under the beta-min condition \citep{buhlmann2011statistics}, this is the variable screening property:
\begin{equation}
    P(\hat{S} \supseteq S_0) \to 1 (p \geq n \to \infty).
\end{equation}
Consistent variable selection is achieved when the irrepresentable condition \citep{zhao2006model} and beta-min condition are satisfied:
\begin{equation}
    P(\hat{S} = S_0) \to 1 (p \geq n \to \infty).
\end{equation}
However, the beta-min condition is not always satisfied with real data (i.e., signals may be weaker than some thresholds). Thus, some signals can be ignored by variable selection in penalized regression methods. Accordingly, a method that can estimate the number of non-zero variables in a model is needed.

In this article, we propose an estimate of the regularization parameter $\lambda$ based on \textit{p}-values from the multi-split regression method. A second-level significance test, higher criticism \citep{donoho2004higher}, is then used to estimate a lower bound of the proportion of false null hypotheses \citep{meinshausen2006estimating} with confidence $(1-\alpha)$. The estimate of $\lambda$ corresponds to the lower bound of the proportion of false null hypotheses. Thus, the optimal choice of $\lambda$ is determined based on not only the mean square error in cross-validation, but also an estimate of non-zero variables.

In the following sections, we describe our analytical framework, including (1) the generation of dependent \textit{p}-values; (2) the derivation of the lower bound for the proportion of false null hypotheses; and (3) its relation to the choice of regularization parameter. We validate our approach with simulation experiments that compare (1) the variable selection of different penalized regression methods in a multi-split setting under different signal sparsity and strength; (2) the consistency of the proportion estimator under different signal sparsity and strengths; and (3) the variable selection results for weak and sparse signals in high-dimensional data with higher criticism tuned parameters. We compare the performance of our analytic approach to traditional penalized regression methods. Finally, we demonstrate our approach using two datasets that were collected with different high throughput genetic technologies. In the first dataset we evaluate the genetic etiology of (1) lipid traits (low-density lipoprotein (LDL) values) in a cohort of patients with Type 2 Diabetes who were participants in the Action to Control Cardiovascular Risk in Diabetes (ACCORD) clinical trial \citep{action2008effects} and have genome-wide single nucleotide polymorphism (SNP) data collected \citep{marvel2017common, rotroff2018genetic1, shah2016genetic, morieri2018genetic, shah2018modulation, rotroff2018genetic2}. In the second dataset, we evaluate epigenetic markers for association with epigenetic analysis evaluating smoking and pulmonary lung function in Agricultural Lung Health Study. Because of different data structures of SNP and epigenetic markers, we demonstrate the performance of our method in different data types. The particular data sets used we used were chosen because of strong, reproducible signals in the data. The etiology of lipid traits has been extensively studies in large cohorts and meta-analyses, which provides a baseline for the expected signals our method should be able to find \citep{dron2016genetics}. The methylation signature of smoking is also well validated \citep{joehanes2016epigenetic, sikdar2019comparison}.

\section{Methods and Materials}
\label{sec:methods}

\subsection{Higher Criticism Test Statistic and Proportion Estimator for Independent Multiple Tests}

Higher criticism is a multiple testing concept originally mentioned by Tukey. \cite{donoho2004higher} proposed a generalized form of higher criticism. In a scenario with a large number of independent hypothesis tests, the higher criticism statistic performs as a global test statistic on the joint null hypothesis at significance level $\alpha$. This test statistic is generated by standardizing the difference of the observed and expected fraction of significance under the global null. In this setting, there are $p$ independent hypotheses tests, where the $i$-th test statistic $T_i$ follows
\begin{center}
    \begin{align*}
    H_{0,i}:T_i &\sim N(0,1),\\
    H_{a,i}:T_i &\sim N(\mu_i,1),
    \end{align*}
\end{center}
where $\mu_i > 0$. In a special case, assume all the non-zero $\mu_i$ are the same. If $\pi$ is defined as the unknown proportion of false null hypotheses, then a global test can be described as 
\begin{center}
    \begin{align*}
    H_{0}:T_i &\sim N(0,1), 1 \leq i \leq p, (i.i.d),\\
    H_{a}:T_i &\sim (1-\pi)N(0,1)+\pi N(\mu,1), 1 \leq i \leq p, (i.i.d).
    \end{align*}
\end{center}
Here, let the \textit{p}-value of the $i$-th hypothesis test be
\begin{equation}
    p_i=P\{ N(0,1) > T_i \},
\end{equation}
and denote $p_{(i)}$ as the $i$-th sorted p-value in increasing order. Note that the \textit{p}-value follows a uniform distribution Unif$(0,1)$ under the null hypotheses. Then, under the global null, $p_{(i)}$ is an order statistic from Unif$(0,1)$. The higher criticism test statistic is defined as:
\begin{equation}
    \text{HC}(i)=\max_{1 \leq i \leq \alpha p} \frac{\sqrt{p}(\frac{i}{p}-p_{(i)})}{\sqrt{p_{(i)}(1-p_{(i)})}}.
\end{equation}
The higher criticism test statistic tests whether the proportion, $\pi$, is $0$. \cite{meinshausen2006estimating} considered an estimator for the proportion of false null hypotheses among independent multiple hypotheses tests. Based on the empirical distribution of the \textit{p}-values of those tests, a lower $100(1-\alpha)\%$ confidence bound, $\hat{\pi}$, is proposed, such that
\begin{equation} \label{ineq}
    P(\hat{\pi} \leq \pi) \geq 1 - \alpha,
\end{equation}
where $\pi=p^{-1} \sum_{i=1}^p \mathbf{1} \{P_i \sim G_i \}$, $P_i$ is the distribution of the $i$-th \textit{p}-value, $G_i$ is an unknown distribution of the $i$-th \textit{p}-value if the $i$-th null hypothesis is rejected. There is no distribution assumption for \textit{p}-values from false null hypotheses, but it is certainly not Unif$(0,1)$. Thus, the empirical distribution of the \textit{p}-values is denoted as
\begin{equation} \label{cdf}
    F_{p}(t)=p^{-1} \sum_{i=1}^p \mathbf{1} \{ P_i \leq t\} = \pi \hat{G}_{\pi p} (t) + (1-\pi) \hat{U}_{(1-\pi)p} (t),
\end{equation}
where $t \in (0,1)$, and $\hat{U}_{(1-\pi)p}$ and $\hat{G}_{\pi p}$ denote the empirical distributions of the \textit{p}-values from $(1-\pi)p$ null hypotheses and $\pi p$ false null hypotheses. When all \textit{p}-values are from null hypotheses, $\pi=0$, then define
\begin{equation} \label{vp}
    V_p(t)=\sup_{t \in (0,1)} \left\{ \frac{\hat{U}_p(t)-t}{\sqrt{t(1-t)}} \right\}
\end{equation}
and denote $\gamma_{p,\alpha}$ as a bounding sequence of $V_p(t)$ such that (a) $p \gamma_{p,\alpha}$ monotonically increases with $p$, and (b) $P(V_p(t) > \gamma_{p,\alpha}) < \alpha$ for all $p$. The estimator of $\pi$ is defined as:
\begin{equation} \label{estimator}
    \hat{\pi}=\sup_{t \in (0,1)} \frac{F_p(t)-t-\gamma_{p,\alpha}\sqrt{t(1-t)}}{1-t},
\end{equation}
where $\hat{\pi}$ satisfies the inequality in Equation (\ref{ineq}).

\subsection{Proportion Estimator for Dependent Multiple Tests}

With the assumption that all variables, hypotheses tests, and p-values are independent, the estimator is determined with Equation (\ref{estimator}). However, this independent assumption is not always true. Since the estimator is based on the higher criticism, the most straightforward way to handle correlation and dependence is by modifying the higher criticism and the corresponding estimator.

Some versions of modified higher criticism consider covariance structure, among other tests. \cite{hall2010innovated} proposed the innovated higher criticism (iHC), a modified higher criticism that detects sparse signals in correlated noise by paying attention to the correlation structure. Again, consider a $p$-dimensional Gaussian vector of test statistics from hypotheses tests,
\begin{equation}
    T = \mathbf{\mu}_p + Z, Z \sim N(\mathbf{0}, \mathbf{\Sigma}_p),
\end{equation}
where $\mathbf{\mu}$ is a vector, and its $i$-th element $\mu_i = 0$ if the $i$-th null hypothesis is true. Let $\mathbf{\Sigma}_p$ be a positive definite matrix, and denote $U_p$ such that $U_p \mathbf{\Sigma}_p U_p^\top = I_p$. For simplicity, when choosing bandwidth $b_p=1$, applying the standard higher criticism to $U_p T$ is a special case of innovated higher criticism, where
\begin{equation}
    \text{iHC}(i)=\sup_{i:1/p \leq p_{(i)} \leq 1/2} \frac{\sqrt{p}(\frac{i}{p}-p_{(i)})}{\sqrt{p_{(i)}(1-p_{(i)})}}.
\end{equation}
Although innovated higher criticism has better performance than higher criticism under the dependence condition, we did not use it on our estimator for three reasons:
\begin{enumerate}
\item In most cases, $\mathbf{\Sigma}_p$ is unknown and must be estimated at high computational cost;
\item Some of the estimated $\mathbf{\Sigma}_p$ may not be positive definite, so $U_p$ cannot be calculated directly;
\item The cost of Cholesky decomposition is $p^3/3$ flops and increases rapidly with $p$.
\end{enumerate}

Another modified test statistic is generalized higher criticism proposed by \cite{barnett2017generalized}. Unlike innovated higher criticism, generalized higher criticism does not transform the original test statistics. Instead, it estimates the variance of distribution to replace the binomial type denominator of standard higher criticism. Based on Equation (\ref{cdf}), the empirical distribution of \textit{p}-values can be defined using test statistics as
\begin{equation}
    S_p(t)=\sum_{i=1}^p \textbf{1}\{|T_i | \geq t\},
\end{equation}
where $t \geq 0$. Then, $\text{cov}\{S_p (t_i),S_p (t_j)\}$, as well as other estimates of $\text{var}(S_p (t))$, can be directly using Hermite polynomial. Generalized higher criticism is defined as
\begin{equation}
    \text{GHC}(t)=\sup_{t \geq 0} \frac{S_p(t)-2p \Bar{\phi}(t)}{\sqrt{\hat{\text{Var}}(S_p(t))}},
\end{equation}
where $\Bar{\phi}(t)=1-\phi(t)$ is the survival function of the standard normal distribution. Generalized higher criticism can handle correlation structure and signal sparsity, but the calculation complexity of estimation can be a problem.

This raises the question of whether there is any other solution since modifying standard higher criticism always results in more complex computations. \cite{jeng2019efficient} studied the consistency of the estimator using standard higher criticism in \cite{meinshausen2006estimating} in a scenario in which there was block dependence. Block dependence means that there is arbitrary dependence within each block of tests but not between blocks. For a sequence of ordered \textit{p}-values, denote $L$ as the total number of noise (true null hypotheses) variables ranked before the last signal (false null hypotheses) variable, and assume that $L$ is bounded almost surely by a number $\Bar{l}$. Let $p_{(1)}^0, ..., p_{(p-\pi p)}^0$ be the ordered \textit{p}-values from true null hypotheses, and assume that for any $r=1,...,\Bar{l}$,
\begin{equation} \label{signal}
    P(p_{(r)}^0 \leq u | p_{(1)}^a, ..., p_{(\pi p)}^a) \leq c_1 F_r(u),
\end{equation}
where $F_r$ is the left-side probability for independent \textit{p}-values and $c_1 \geq 1$ is a constant. Theorem 2.4 of \cite{jeng2019efficient} stated that under block dependence and Equation (\ref{signal}), let the true but unknown $\pi = p^{-\eta}$ and $\eta \in [0,1)$, then we have $P(1-\delta < \hat{\pi}/\pi <1) \to 1$ as $p \to \infty$ for an arbitrary small constant $\delta > 0$ if any of these conditions below are satisfied:
\begin{enumerate}
\item $\eta \in [0,(1-\kappa)/2)$, inf$_{t \in (0,1)}G'(t)=0$, and $1 << \Bar{l}\log(\Bar{l}) << p^{1-\eta/2}$.
\item $\eta \in [(1-\kappa)/2,2/3)$, $G(p^{-\tau}) \to 1$ for some $\tau > 2 \eta - (1-\kappa)$, and $1 << \Bar{l}\log(\Bar{l}) << p^{1-\eta/2}$.
\item $\eta \in [2/3,1)$, $\kappa \in [0,1)$, $G(p^{-\tau}) \to 1$ for some $\tau > 2 \eta - (1-\kappa)$, and $1 << \Bar{l}\log(\Bar{l}) << p^{2(1-\eta)}$.
\end{enumerate}
As these conditions are realistic and the consistency of the estimator from standard higher criticism is proven, we use a standard higher criticism estimator in our study.

\subsection{\textit{p}-values in High-dimensional Regression}

Many prior works have focused on calculating \textit{p}-values in high-dimension regression models. We discuss here a few that are related to our work. Before considering details, we first define our notations in a linear regression model:
\begin{equation} \label{linear}
    \mathbf{y}=\mathbf{X}\mathbf{\beta} + \mathbf{\epsilon},
\end{equation}
where $\mathbf{y}$ is an $n \times 1$ vector, $\mathbf{X}$ is an $n \times p$ design matrix and $p>>n$, $\beta$ is a $p \times 1$ parameter vector, and error term $\epsilon$ is an $n \times 1$ vector. To eliminate the intercept in the regression model, the observed variable and input variable are centered so that the observed mean is 0. \cite{tibshirani1996regression} defined the Lasso estimator, one of the most well-known ways to perform regression shrinkage and variable selection, as:
\begin{equation}
    \hat{\mathbf{\beta}}_{p \times 1}^{\text{Lasso}} (\lambda) = \argminB_{\mathbf{\beta}} \{ || \mathbf{y} - \mathbf{X} \mathbf{\beta} ||_2^2 + \lambda || \mathbf{\beta} ||_1\}, 
\end{equation}
where $||\cdot||_1$ is vector $L1$-norm, and $\lambda \geq 0$ is the tuning parameter. There are three main inferential tasks in high-dimensional models \cite{buhlmann2011statistics}: given that $\beta^0$ is the truth, (a) prediction accuracy, $||\mathbf{X} \hat{\mathbf{\beta}}-\mathbf{X} \mathbf{\beta}^0||_2^2/n$; (b) parameter estimation accuracy, $||\hat{\mathbf{\beta}}-\mathbf{\beta}^0||_q$, $q \in \{1,2\}$; and (c) variable selection by considering $\hat{S}$ and $S^0$, where $\hat{S}=\{ i; \hat{\beta}_i \neq 0, i=1,...,p \}$ and$S_0=\{ i; \beta_i^0 \neq 0, i=1,...,p \}$. 

A number of prior studies have focused on assigning significance in high-dimensional regression. This is a challenging task since asymptotically valid \textit{p}-values are not available. \cite{meinshausen2009p} proposed randomly splitting samples into two groups multiple times, say $B$. The authors conducted variable selection on one subgroup of samples to choose $k_{i=1,...,B} < n << p$ variables and performed ordinary least squares for $k_i$ variables on the other subgroup of samples. Subsequently, $B$ groups of \textit{p}-values were aggregated for all variables. 

Post-selection inference is another widely accepted method of assigning significance in high-dimensional regression. \cite{lockhart2014significance} originally described the method and \cite{tibshirani2016exact} extended the results of the asymptotic null distribution of the "covariance" test to test hypotheses up on adding new individual features to a selected model.

In the present study, we split the data multiple times. We apply Lasso, adaptive Lasso \citep{zou2006adaptive}, smoothly clipped absolute deviation (SCAD) \citep{fan2001variable}, and minimax concave penalty (MCP) \citep{zhang2010nearly} to the first half of the data to perform variable selection. We chose these methods, because first, we would like to compare their penalties in regression; second, they satisfy the screening property when assuming the compatibility condition on the design matrix $\mathbf{X}$, the sparsity assumption, and a beta-min condition \citep{dezeure2015high}. We use the ordinary least squares method for the second half of the data to obtain corresponding \textit{p}-values for the individual variables selected from the first half of the data. After sufficient Monte Carlo sampling, we aggregate the \textit{p}-values for each individual variable. While this is a generic method, it requires a beta-min assumption. Details of our proposed method are described in Section \ref{hctr}.

\subsection{Higher Criticism Tuned Regression Algorithm} \label{hctr}

\begin{figure}[ht]
  \centering
  \includegraphics[width=1\textwidth]{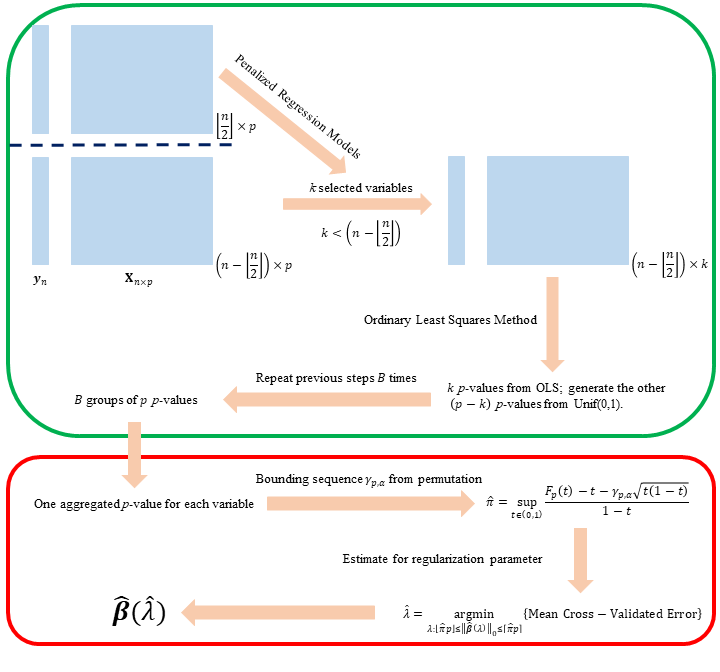}
  \caption[Flowchart of higher criticism tuned regression]{Flowchart of higher criticism tuned regression. \textit{p}-value generation is in the green circle, and the estimated regression solution is in the red circle.}
  \label{fig:fig1}
\end{figure}

To perform high-dimensional variable selection, we propose a robust multiple-stage algorithm called higher criticism tuned regression. Figure \ref{fig:fig1} is a flowchart that outlines details of the procedure. The algorithm contains two main parts: (a) \textit{p}-value generation (in the green circle) and (b) a corresponding estimated lower bound $\hat{\pi}$ and higher criticism tuned regression solution $\hat{\beta}^{\text{HCR}}(\lambda)$ (in the red circle).

In the first part, \textit{p}-values for each variable in the high-dimension model are generated according to single-split \citep{wasserman2009high} and multi-split \citep{meinshausen2009p} methods. Recall that Equation (\ref{linear}) is a linear regression model, where $\mathbf{y}$ is an $n \times 1$ vector, and $\mathbf{X}$ is an $n \times p$ design matrix. We randomly choose $\floor*{\frac{n}{2}}$ observations from $\mathbf{y}$ and their corresponding rows in $\mathbf{X}$. We apply penalized regression methods, such as Lasso, adaptive Lasso, SCAD, or MCP, on the first half of data, $\mathbf{y}_{\floor*{\frac{n}{2}}}$ and $\mathbf{X}_{\floor*{\frac{n}{2}} \times p}$. Assume that we have $k$ non-zero estimates out of $p$ variables in penalized regression, where $k<(n-{\floor*{\frac{n}{2}}})$. This creates a subset of the second half of $\mathbf{X}$, and so now we have $\mathbf{X}_{ (n-{\floor*{\frac{n}{2}}}) \times k}$. The ordinary least squares method is applied for $\mathbf{y}_{n-{\floor*{\frac{n}{2}}}}$ and $\mathbf{X}_{ (n-{\floor*{\frac{n}{2}}}) \times k}$. To control the family-wise error rate, the $i$-th \textit{p}-value from \textit{t}-test is corrected by Bonferroni correction as $p_i=\min(kp_i,1)$. However, as mentioned in \cite{meinshausen2006estimating}, using the family-wise error rate control results in a significant loss of power. If the signals are weak and the proportion of false null hypotheses is small, a more powerful tool is needed instead of stricter inference. Further, we empirically calculate the distribution of \textit{p}-values under the null hypothesis and use this for subsequent estimator construction. We are interested in the difference between the null distribution and the observed values and thus we do not use family-wise error rate control for null distribution \textit{p}-values or sample calculated \textit{p}-values. We assign the $k$ \textit{p}-values from the \textit{t}-test in ordinary least squares regression to the corresponding $k$ selected variables. In contrast to \cite{meinshausen2009p}, we denote the \textit{p}-values of the rest $(p-k)$ unselected variables as random variables following uniform distribution between $(0,1)$. We repeat random sample splitting and calculate the \textit{p}-values $B$ times, where $B$ should be sufficiently large. For example, if $B=100$, we will have $B$ \textit{p}-values for each variable, $p_i^1,...,p_i^B,(i=1,...,p)$. These $B$ \textit{p}-values are aggregated by harmonic mean \textit{p}-value \cite{wilson2019harmonic}. Similar to Fisher’s method, the harmonic mean \textit{p}-value is a \textit{p}-value used to test whether groups of \textit{p}-values are statistically significant. Because the half samples used in ordinary least squares regression are part of the same full sample, there is dependence among the $B$ \textit{p}-values. The advantage of the harmonic mean \textit{p}-value over Fisher’s method is that independence among \textit{p}-values is not required. 

After generating \textit{p}-values for all variables, we use them to build $\hat{\pi}=\sup_{t \in (0,1)} \frac{F_p(t)-t-\gamma_{p,\alpha}\sqrt{t(1-t)}}{1-t}$ in equation \ref{estimator}. First, recall that $F_p (t)$ is the empirical distribution of $p$ \textit{p}-values from test data. Then, to get a valid bounding sequence $\gamma_{p,\alpha}$, \cite{jeng2019efficient} simulated the empirical distribtion of $V_p (t)$ in equation \ref{vp} under global null, where the null hypothesis is realized by using permutation method proposed in \cite{westfall1993resampling}. As a result, $\gamma_{p,\alpha}$ can be determined by finding the $(1-\alpha_p)$th quantile of the empirical distribution of $V_p (t)$ from simulations. 

Previously, the choice of tuning parameter $\lambda$ was always data-driven, for example, loss for linear regression or likelihood and misclassification rate for logistic regression. Here, based on $\hat{\pi}$, we propose a narrower tuning region for $\lambda$ in penalized regression methods:
\begin{equation}
    \hat{\lambda}=\argminB_{\lambda:\floor*{\hat{\pi}p} \leq || \hat{\mathbf{\beta}}(\lambda) ||_0 \leq \ceil*{\hat{\pi}p}} \left\{ \text{Mean Cross-Validated Error} \right\},
\end{equation}
where $p$ is the total number of variables. The widely accepted tuning region of $\lambda$ is $[0,\lambda_0]$, where $\lambda_0$ is the minimal value that shrinks all variables to zero. Since $\pi \geq 0$ and $\hat{\pi} \geq 0$, we have $\hat{\lambda}(\hat{\pi}p) \leq \lambda_0$. Thus, $[\hat{\lambda}(\floor*{\hat{\pi}p}), \hat{\lambda}(\ceil*{\hat{\pi}p})] \subseteq [0, \lambda_0]$ gives a narrower tuning region of $\lambda$ and a better final choice of $\lambda$ if $\pi > 0$.
\begin{figure}
  \centering
  \includegraphics[width=0.618\textwidth]{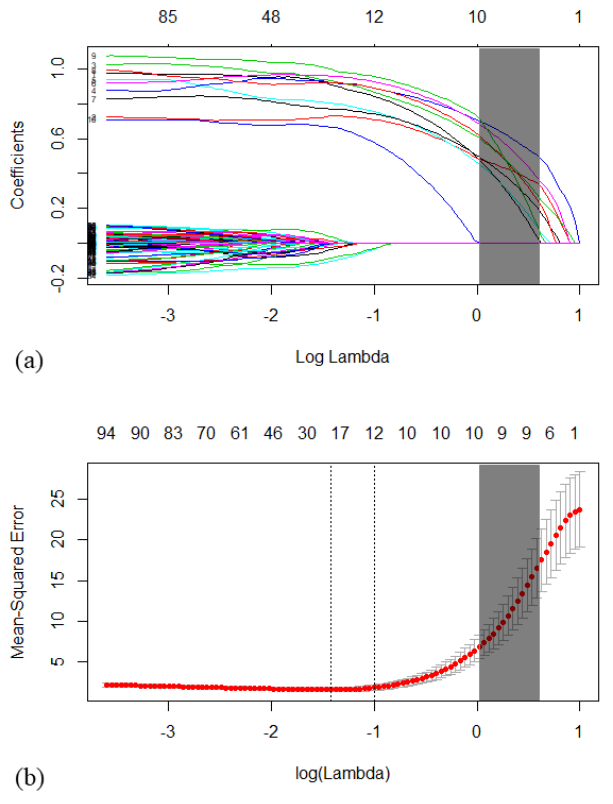}
  \caption[An example of a Lasso (a) solution path and (b) cross-validation curve]{An example of a Lasso (a) solution path and (b) cross-validation curve (drawn with  \textit{glmnet}). An updated tuning region (in gray) for the regularization parameter in Lasso is suggested by an estimated value of non-zero variable proportion. }
  \label{fig:fig2}
\end{figure}
\FloatBarrier

For example, Figure \ref{fig:fig2} shows an example with a Lasso solution path using \textit{glmnet}. The general tuning method for the regularization parameter is finding a minimum $\lambda$ such that all the estimates are shrunk to zeros. The value of $\lambda$ that gives the minimum mean cross-validated error is $0.242$, which gives $\log(\lambda)=-1.419$, the left dashed line in Figure \ref{fig:fig2}(b). Assume we have $\hat{\pi}p=9.5$. Zooming into Figure \ref{fig:fig2}(a), the gray area is the new tuning region, which is bounded by $\{\lambda_1 ; ||\hat{\mathbf{\beta}}(\lambda) ||_0 =9\}$ and $\{\lambda_2 ; ||\hat{\mathbf{\beta}}(\lambda) ||_0 =10\}$. When the new tuning region is shifted to Figure \ref{fig:fig2}(b), we can find a new $\hat{\lambda}$ that minimizes cross-validated errors. In the following section, we demonstrate how this effects our final selected model.

We use simulation tests to answer the following topics: (a) whether a combination of different methods will perform better than a single method for multi-split variable selection; (b) the performance of \textit{p}-value aggregation; (c) the consistency of the lower bound estimator; and (d) the accuracy of variable selection using higher criticism tuned regression.

\section{Simulation Studies}
\label{sec:simulation}

\subsection{Variable Selection with Different Signal Strengths under Block Dependence} \label{diff_strength}

We first consider sparse linear and logistic regression models with different signal strengths and compare the variable selection performance of penalized regression methods: Lasso, adaptive Lasso, SCAD, and MCP. For the simulation studies, we simulate $n=100$ samples and $p=1000$ variables. Each row of design matrix $X_{n \times p}$ is generated as $x_i \sim N_p (\mathbf{0}, \mathbf{\Sigma})$ with block dependence structure matrix $\mathbf{\Sigma}=\text{diag}(\Sigma_{1}, \Sigma_{2}, \Sigma_{3}, ...)$, where each $\Sigma_{i}$ is a $\pi p \times \pi p$ matrix with element $\sigma_{j,k} =(0.5^{|j-k|})$. Note that our simulations focus on dependent multiple tests using a block correlation structure, which is a more realistic and challenging scenario than an independent variable design. To implement the multi-split algorithm, we conduct penalized regression in $B=100$ loops. In each loop, $50$ samples are randomly chosen from a total $100$ samples for penalized regression. Then, $5$-fold cross-validation tests are applied for Lasso, adaptive Lasso, SCAD, and MCP to search for the corresponding best model using two R packages \textit{glmnet} \citep{friedman2009glmnet} and \textit{ncvreg} \citep{breheny2019package}. The parameter $\gamma$ is fixed at $3.7$ in SCAD and $3.0$ in MCP. Tables \ref{table_1} and \ref{table_2} summarize the average true positive (TP, a variable in the true model is chosen correctly), false positive (FP), true negative (TN), and false negative (FN) values. Besides, diagnostic odds ratio \citep{glas2003diagnostic} and $\text{F}_1$ score \citep{chinchor1992muc} are calculated for each method, where 
\begin{equation} \label{dor}
    \text{DOR} = \frac{\text{TP}\cdot\text{TN}}{\text{FP}\cdot\text{FN}},
\end{equation}
\begin{equation} \label{f1}
    \text{F}_1 = \left ( \frac{2}{\text{recall}^{-1} + \text{precision}^{-1}} \right ) = 
    \frac{2\text{TP}}{2\text{TP} + \text{FP} + \text{FN}}.
\end{equation}
From Equation \ref{dor} and \ref{f1}, we can see $\text{F}_1$ score focuses more on sensitivity of a test, while diagnostic odds ratio is a more general test. It is possible that they do not give the same ranking of tests.

\begin{table}
    \caption{Comparison of methods for different signal strengths in linear models}
    \centering
    \begin{tabular}{cccccccc}
    Methods & $\beta_{i} \neq 0, \text{where }i=1,...,10$ & TP & FP & TN & FN & DOR & $\text{F}_1$ score \\
    \midrule\midrule
    \multirow{4}{*}{Lasso} & 0.2 & 2.51 & 9.09 & 980.91 & 7.49 & 36.162 & 0.232 \\
    & 0.5 & 8.06 & 19.50 & 970.50 & 1.94 & 206.773 & 0.429 \\
    & 1 & 9.83 & 25.80 & 964.20 & 0.17 & 2160.986 & 0.431 \\
    & 2 & 9.97 & 27.05 & 962.95 & 0.03 & 11830.7 & 0.424 \\
    \hline
    \multirow{4}{*}{Adaptive Lasso} & 0.2 & 4.46 & 36.05 & 953.95 & 5.54 & 21.303 & 0.177 \\
    & 0.5 & 8.51 & 27.31 & 962.69 & 1.49 & 201.330 & 0.371 \\
    & 1 & 9.78 & 23.16 & 966.84 & 0.22 & 1855.805 & 0.456 \\
    & 2 & 9.87 & 17.61 & 972.39 & 0.13 & 4192.325 & 0.527 \\
    \hline
    \multirow{4}{*}{SCAD $(\gamma = 3.7)$} & 0.2 & 2.17 & 6.28 & 983.72 & 7.83 & 43.412 & 0.235 \\
    & 0.5 & 5.80 & 10.25 & 979.75 & 4.20 & 131.999 & 0.445 \\
    & 1 & 6.17 & 9.99 & 980.01 & 3.83 & 158.034 & 0.472 \\
    & 2 & 6.50 & 9.80 & 980.20 & 3.50 & 185.752 & 0.494 \\
    \hline
    \multirow{4}{*}{MCP $(\gamma = 3)$} & 0.2 & 1.08 & 1.94 & 988.06 & 8.92 & 61.665 & 0.166 \\
    & 0.5 & 3.55 & 2.76 & 987.24 & 6.45 & 196.871 & 0.435 \\
    & 1 & 4.38 & 3.96 & 986.04 & 5.62 & 194.061 & 0.478 \\
    & 2 & 4.83 & 3.55 & 986.45 & 5.17 & 259.599 & 0.526 \\
    \midrule\midrule
    \end{tabular}
    \label{table_1}
\end{table}

\begin{table}
    \caption{Comparison of methods for different signal strengths in logistic models}
    \centering
    \begin{tabular}{cccccccc}
    Methods & $\beta_{i} \neq 0, \text{where }i=1,...,10$ & TP & FP & TN & FN & DOR & $\text{F}_1$ score \\
    \midrule\midrule
    \multirow{4}{*}{Lasso} & 0.2 & 0.43 & 13.15 & 976.85 & 9.57 & 3.338 & 0.036 \\
    & 0.5 & 1.82 & 11.41 & 978.59 & 8.18 & 19.082 & 0.157 \\
    & 1 & 3.39 & 13.55 & 976.45 & 6.61 & 36.958 & 0.252 \\
    & 2 & 3.33 & 10.69 & 979.31 & 6.67 & 45.736 & 0.277 \\
    \hline
    \multirow{4}{*}{Adaptive Lasso} & 0.2 & 0.47 & 21.35 & 968.65 & 9.53 & 2.238 & 0.030 \\
    & 0.5 & 2.72 & 18.93 & 971.07 & 7.28 & 19.166 & 0.172 \\
    & 1 & 3.86 & 15.79 & 974.21 & 6.14 & 38.787 & 0.260 \\
    & 2 & 4.28 & 13.76 & 976.24 & 5.72 & 53.087 & 0.305 \\
    \hline
    \multirow{4}{*}{SCAD $(\gamma = 3.7)$} & 0.2 & 0.11 & 3.42 & 986.58 & 9.89 & 3.209 & 0.016 \\
    & 0.5 & 1.28 & 4.99 & 985.01 & 8.72 & 28.976 & 0.157 \\
    & 1 & 2.51 & 7.60 & 982.40 & 7.49 & 43.318 & 0.250 \\
    & 2 & 3.09 & 8.77 & 981.23 & 6.91 & 50.032 & 0.283 \\
    \hline
    \multirow{4}{*}{MCP $(\gamma = 3)$} & 0.2 & 0.02 & 1.56 & 988.44 & 9.98 & 1.270 & 0.003 \\
    & 0.5 & 0.64 & 1.46 & 988.54 & 9.36 & 46.296 & 0.106 \\
    & 1 & 1.48 & 2.28 & 987.72 & 8.52 & 75.253 & 0.215 \\
    & 2 & 1.95 & 2.43 & 987.57 & 8.05 & 98.447 & 0.271 \\
    \midrule\midrule
    \end{tabular}
    \label{table_2}
\end{table}

In sparse linear regression, the response is $\mathbf{y}=\mathbf{X}\mathbf{\beta} + \mathbf{\epsilon}$, where $\mathbf{\beta}=(\mathbf{\beta}_{\pi p}^\top, \mathbf{0}_{p-\pi p}^\top)^\top$, $\beta_{\pi p}$ is a vector with equal elements, and $\pi$ is fixed at $0.01$, $\epsilon \sim N_p(\mathbf{0}, \mathbf{I}_p) $. For all four methods, more signals are chosen as the signal strengths increases. Among the four methods, Lasso chooses more false positive signals when the signal is strong, while adaptive Lasso results in less false positive values when the signal is strong. However, both Lasso and adaptive Lasso have lower specificity than SCAD and MCP. The diagnostic odds ratio (DOR) and $\text{F}_1$ score are more general indicators of test performance. We say a completely random test will have DOR $=1$, so a larger DOR indicates a more significant classification test. Both DOR and $\text{F}_1$ score show us the fact that it is more challenging to identify signals when their strengths are weak. Table \ref{table_1} indicates two points that could be improved: (1) the power of SCAD and MCP, as they are less likely to produce type I errors than the other two methods; and (2) the power of all methods when signals are weak. We later compare our proposed methods with the results.

Note that in a genomics situation, detection of any one variable, such as single-nucleotide polymorphism (SNP), in a block would usually be sufficient to trigger detailed fine mapping to understand the inter-variable correlation or linkage disequilibrium (LD) in genetics. In a word, if one of variables in a block is detected, then you have effectively detected all correlated variables. The power of this method is underestimated if this is a genetic simulation.

In sparse logistic regression, each element in response $\mathbf{y}$ follows a Bernoulli distribution with the success probability as $\frac{\exp(\mathbf{x}_i \mathbf{\beta})}{(1+\exp(\mathbf{x}_i \mathbf{\beta}))}$, where $\mathbf{x}_i$ is the $i$-th row of design matrix $\mathbf{X}$ and $\mathbf{\beta}$ is the same as denoted in the linear regression models. Local solutions of SCAD and MCP are implemented in logistic regression models. Table \ref{table_2} shows that it is more challenging to find the true model in logistic regression models than in linear regression models. The results also indicate that power can be improved when signals are weak. This is a high-dimensional classification problem and is the focus of other projects.

\subsection{Variable Selection with Different Signal Sparsity under Block Dependence} \label{diff_sparsity}

We now look at variable selection performance of the four methods at different sparsity values, $\pi=0.001$, $0.002$, $0.005$, $0.01$. The setup for all models is similar to that outlined in Section 3.1. Note that all row vectors in the design matrix are generated randomly following multivariate normal distribution, so increasing the number of variables in the true model may not increase the number in the final chosen model for some algorithms.

Again, we first fit linear regression models. Table \ref{table_3} shows that at all sparsity levels, adaptive Lasso is always the most powerful method, followed by Lasso. However, these models result in more false positive values than SCAD and MCP. Among all the methods, MCP has the highest specificity. Table \ref{table_4} summarizes the logistic regression models and shows the same conclusions as Table \ref{table_3}. Now we revisit the penalties of these four penalized regression methods and the reason why we choose them for comparison. Lasso is used as a baseline here since it is the most basic shrinkage regression methods while generating biased estimators. In an orthonormal case, we have 
\begin{equation}
    \left\{
	    \begin{array}{lr}
            E|\hat{\beta}_j-\beta_j| = 0, & \beta_j = 0; \\
            E|\hat{\beta}_j-\beta_j| \approx \beta_j, & 0 < |\beta_j| \leq \lambda; \\
            E|\hat{\beta}_j-\beta_j| \approx \lambda, & |\beta_j| > \lambda,
	    \end{array}
    \right.
\end{equation}
where $\hat{\beta}_j$ is the $j$-th Lasso estimator, and $\beta_j$ is the corresponding true value. Based on Lasso, Adaptive Lasso has a two-stage algorithm. In order to reduce bias, Adaptive Lasso assigns smaller penalties to variables with larger initial regression coefficients from OLS or ridge regression. As alternatives to Adaptive Lasso, some single-stage algorithms have been proposed where the penalty tapers off when $|\beta_j|$ gets larger. Tapering penalties cannot be convex. Famous examples of them are SCAD and MCP, both with nonconvex penalties. This is where the $\gamma$ in the tables from, which controls how rapidly the penalty tapers off.

\begin{table}
    \caption[Comparison of methods for different signal sparsity in linear models]{Comparison of methods for different signal sparsity in linear models, $p=1000$ and $\beta_{i}=0.5$, where $i=1,...,\pi p$}
    \centering
    \begin{tabular}{ccccccccc}
    $\pi p$ & Methods & TP & FP & TN & FN & DOR & $\text{F}_1$ score \\
    \midrule\midrule
    \multirow{4}{*}{1} & Lasso & 0.65 & 9.06 & 989.94 & 0.35 & 202.921 & 0.121 \\
    & Adaptive Lasso & 0.94 & 39.60 & 959.40 & 0.06 & 379.561 & 0.045 \\
    & SCAD $(\gamma = 3.7)$ & 0.58 & 6.99 & 992.01 & 0.42 & 195.983 & 0.135 \\
    & MCP $(\gamma = 3)$ & 0.49 & 1.98 & 997.02 & 0.51 & 483.799 & 0.282 \\
    \hline
    \multirow{4}{*}{2} & Lasso & 1.91 & 10.08 & 987.92 & 0.09 & 2079.946 & 0.273 \\
    & Adaptive Lasso & 1.94 & 35.16 & 962.84 & 0.06 & 885.433 & 0.099 \\
    & SCAD $(\gamma = 3.7)$ & 1.74 & 7.71 & 990.29 & 0.26 & 859.575 & 0.304 \\
    & MCP $(\gamma = 3)$ & 1.32 & 2.05 & 995.95 & 0.68 & 943.080 & 0.492 \\
    \hline
    \multirow{4}{*}{5} & Lasso & 4.49 & 15.09 & 979.91 & 0.51 & 571.706 & 0.365 \\
    & Adaptive Lasso & 4.63 & 29.41 & 965.59 & 0.37 & 410.844 & 0.237 \\
    & SCAD $(\gamma = 3.7)$ & 3.81 & 7.69 & 987.31 & 1.19 & 411.060 & 0.462 \\
    & MCP $(\gamma = 3)$ & 2.67 & 2.05 & 992.95 & 2.33 & 555.046 & 0.549 \\
    \hline
    \multirow{4}{*}{10} & Lasso & 8.06 & 19.50 & 970.50 & 1.94 & 206.773 & 0.429 \\
    & Adaptive Lasso & 8.51 & 27.31 & 962.69 & 1.49 & 201.330 & 0.371 \\
    & SCAD $(\gamma = 3.7)$ & 5.80 & 10.25 & 979.75 & 4.20 & 131.999 & 0.445 \\
    & MCP $(\gamma = 3)$ & 3.55 & 2.76 & 987.24 & 6.45 & 196.871 & 0.435 \\
    \midrule\midrule
    \end{tabular}
    \label{table_3}
\end{table}

\begin{table}
    \caption[Comparison of methods for different signal sparsity in logistic models]{Comparison of methods for different signal sparsity in logistic models, $p=1000$ and $\beta_{i}=1$, where $i=1,...,\pi p$}
    \centering
    \begin{tabular}{ccccccccc}
    $\pi p$ & Methods & TP & FP & TN & FN & DOR & $\text{F}_1$ score \\
    \midrule\midrule
    \multirow{4}{*}{1} & Lasso & 0.32 & 10.70 & 988.30 & 0.68 & 43.466 & 0.053 \\
    & Adaptive Lasso & 0.43 & 22.43 & 976.57 & 0.57 & 32.845 & 0.036 \\
    & SCAD $(\gamma = 3.7)$ & 0.13 & 2.16 & 996.84 & 0.87 & 68.960 & 0.079 \\
    & MCP $(\gamma = 3)$ & 0.08 & 1.26 & 997.74 & 0.92 & 68.857 & 0.068 \\
    \hline
    \multirow{4}{*}{2} & Lasso & 1.52 & 10.01 & 987.99 & 0.48 & 312.551 & 0.225 \\
    & Adaptive Lasso & 1.79 & 15.71 & 982.29 & 0.21 & 532.963 & 0.184 \\
    & SCAD $(\gamma = 3.7)$ & 1.62 & 7.29 & 990.71 & 0.38 & 579.363 & 0.297 \\
    & MCP $(\gamma = 3)$ & 1.30 & 2.28 & 995.72 & 0.70 & 811.050 & 0.466 \\
    \hline
    \multirow{4}{*}{5} & Lasso & 2.50 & 9.72 & 985.28 & 2.50 & 101.366 & 0.290 \\
    & Adaptive Lasso & 2.80 & 12.32 & 982.68 & 2.20 & 101.517 & 0.278 \\
    & SCAD $(\gamma = 3.7)$ & 2.70 & 10.66 & 984.34 & 2.30 & 108.399 & 0.294 \\
    & MCP $(\gamma = 3)$ & 1.80 & 3.16 & 991.84 & 3.20 & 176.554 & 0.361 \\
    \hline
    \multirow{4}{*}{10} & Lasso & 3.39 & 13.55 & 976.45 & 6.61 & 36.958 & 0.252 \\
    & Adaptive Lasso & 4.12 & 16.20 & 973.80 & 5.88 & 42.119 & 0.272 \\
    & SCAD $(\gamma = 3.7)$ & 2.61 & 7.60 & 982.40 & 7.39 & 45.653 & 0.258 \\
    & MCP $(\gamma = 3)$ & 1.35 & 1.91 & 988.09 & 8.65 & 80.739 & 0.204 \\
    \midrule\midrule
    \end{tabular}
    \label{table_4}
\end{table}

\subsection{\textit{p}-value Aggregation} \label{p_aggre}

\begin{figure}
  \centering
  \includegraphics[width=0.6\textwidth]{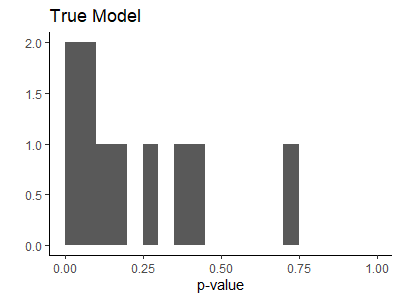}
  \caption[Distributions of all \textit{p}-values from multiple linear regression of the true model]{Distributions of all \textit{p}-values from multiple linear regression of the true model, where the proportion of signal is $0.01$, $p=1000$, the signal strength $\beta_i=0.2$.}
  \label{fig:fig3}
\end{figure}

After conducting variable selection on the first-half samples with penalized regression models (see Section 3.1 and 3.2), we use classic ordinary least squares regression to calculate \textit{p}-values for those selected variables and assign Unif$(0,1)$ to the unselected variables. After generating $B$ groups of \textit{p}-values, we implement \textit{harmonicmeanp}, as described in \cite{wilson2019harmonic}, to calculate the corresponding asymptotically exact harmonic mean \textit{p}-values. For comparison, we fit the true model with all signals and plot the distribution of \textit{p}-values from the fit in Figure \ref{fig:fig3}, where, for model $\beta_i=0.2$, $\sigma^2=1$ and $i$ is the index of signals.
Because of multicollinearity and the weakness of signals, the distribution of \textit{p}-values is highly uniform.
\begin{figure}
  \centering
  \includegraphics[width=1\textwidth]{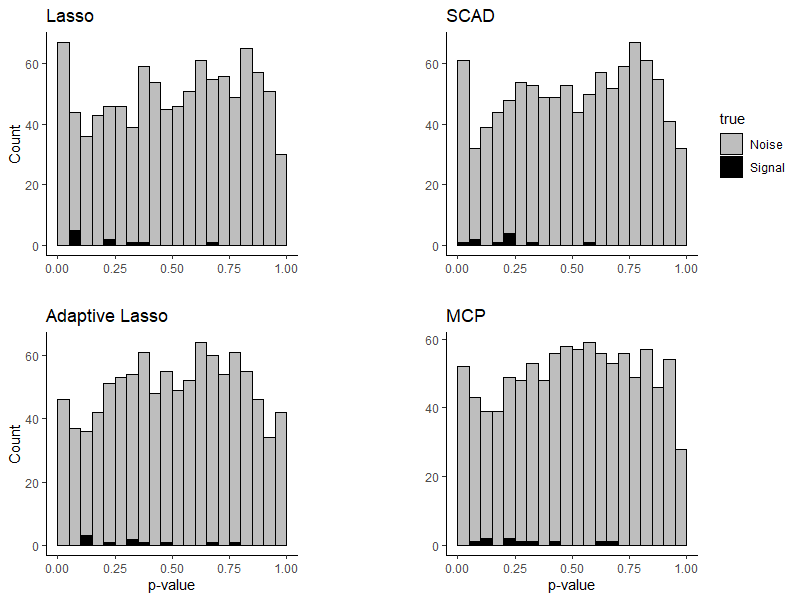}
  \caption[Distributions of all \textit{p}-values from multiple linear regression of weak signal detection]{Distributions of all \textit{p}-values from multiple linear regression of weak signal detection using $4$ penalized regression methods, where proportion of signal is $0.01$, $p=1000$, and signal strength $\beta_i=0.2$.}
  \label{fig:fig4}
\end{figure}
\begin{figure}
  \centering
  \includegraphics[width=1\textwidth]{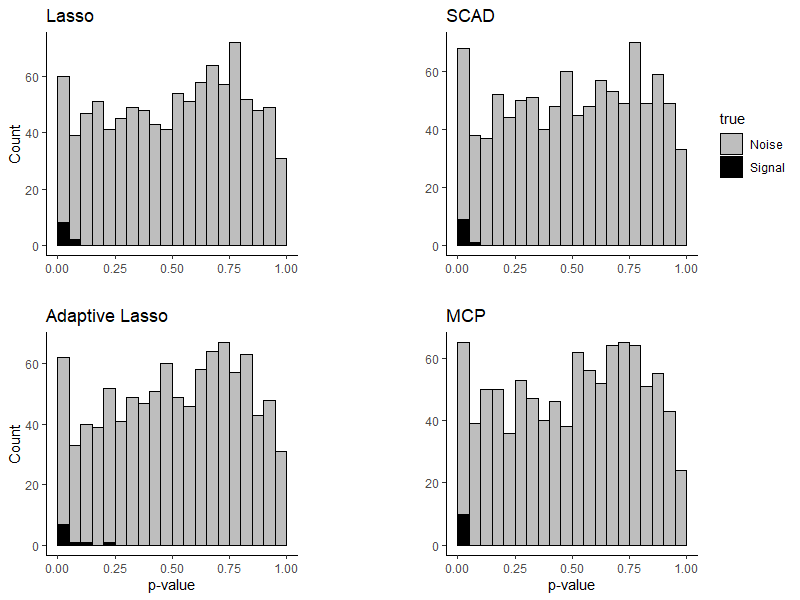}
  \caption[Distributions of all \textit{p}-values from multiple linear regression of moderate signal detection]{Distributions of all \textit{p}-values from multiple linear regression of moderate signal detection using $4$ penalized regression methods, where the proportion of signal is $0.01$, $p=1000$, and signal strength $\beta_i=0.5$.}
  \label{fig:fig5}
\end{figure}
\begin{figure}
  \centering
  \includegraphics[width=1\textwidth]{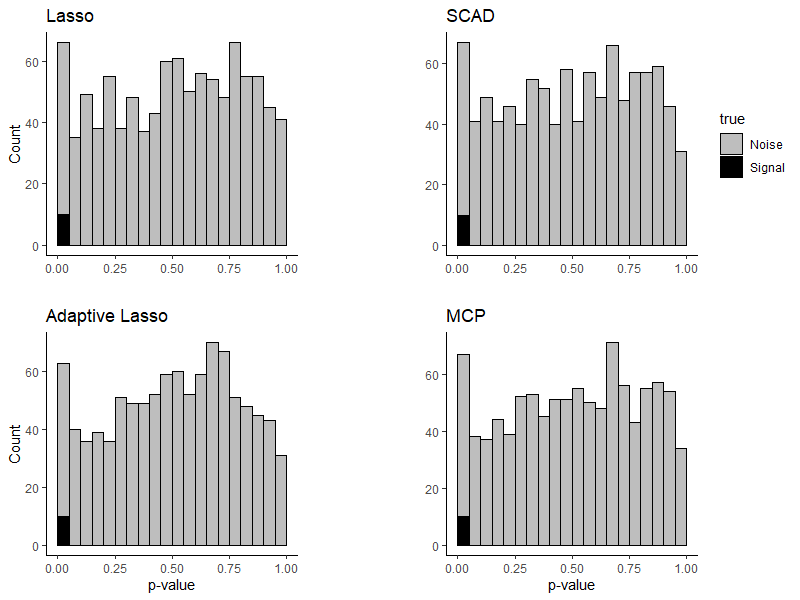}
  \caption[Distributions of all \textit{p}-values from multiple linear regression of strong signal detection]{Distributions of all \textit{p}-values from multiple linear regression of strong signal detection using $4$ penalized regression methods, where the proportion of signal is $0.01$, $p=1000$, and signal strength $\beta_i=1$.}
  \label{fig:fig6}
\end{figure}
Figure \ref{fig:fig4}, \ref{fig:fig5}, and \ref{fig:fig6} show the distributions of the aggregated \textit{p}-values, with \textit{p}-values of signals in black and \textit{p}-values of noise in gray. When the signal is weak, such as in Figure \ref{fig:fig4}, $\beta_i=0.2$, where $i$ is the location of a signal, and not all signals have \textit{p}-values smaller than $0.05$. This indicates that using \textit{p}-values for variable selection is not desirable and that the beta-min condition must be satisfied \cite{buhlmann2011statistics}. However, as signal strength increases to $0.5$ and $1$, the \textit{p}-values of signals become increasingly significant. Further, although there is a clear difference in variable selection performance in Section \ref{diff_strength} and \ref{diff_sparsity}, the distributions of the \textit{p}-values of signals are not significantly different. Finally, we conclude that our modified multi-split algorithm works better for strong signals. We use the permutation method to repeat this procedure and generate a valid higher criticism test statistic under the null hypothesis.

\subsection{Lower Bound Estimation Performance}

Using the same models in Section \ref{diff_strength} through \ref{p_aggre}, we follow the permutation method introduced in \cite{jeng2019efficient} to shuffle outcomes $\mathbf{y}$ and re-assign each element of $\mathbf{y}$ to a random individual. For each permutated data set, we calculate a new group of \textit{p}-values and a corresponding higher criticism test statistic. We repeat this process $1,000$ times and construct an empirical distribution of higher criticism test statistic under the null hypothesis. As suggested in \cite{jeng2019efficient}, we use $\alpha = 1/\sqrt{\log(p)}$ for our test, where $p=1000$. Using Equation (\ref{estimator}), we calculate an estimate of signal proportion, $\hat{\pi}$, and compare it with the true signal proportion, $\pi$. 
\begin{table}
    \caption[Mean and standard deviation of proportion estimates over true proportion] {Eight groups of $\hat{\pi}/\pi$ (SE), mean and standard error of mean for \textit{proportion estimates over true proportion}, where $\pi$ is the true signal proportion, $\beta_{i} \neq 0$ is the signal strength, and the sample size is $10$.}
    \centering
    \begin{tabular}{ccccccc}
    $\pi$ & $\beta_{i}$ & Lasso & Adaptive Lasso & SCAD $(\gamma = 3.7)$ & MCP $(\gamma = 3)$ & \textbf{Combination} \\
    \midrule\midrule
    \multirow{2}{*}{0.01} & 0.5 & 0.425 (0.150) & 0.318 (0.114) & 0.796 (0.101) & 0.872 (0.079) & \textbf{0.563 (0.072)} \\
    & 1 & 1.142 (0.143) & 0.952 (0.057) & 1.026 (0.036) & 1.230 (0.213) & \textbf{0.994 (0.031)} \\
    \hline
    & 0.2 & 0.149 (0.074) & 0.402 (0.402) & 0.583 (0.280) & 0.310 (0.251) & \textbf{0.185 (0.124)} \\
    0.005 & 0.5 & 0.831 (0.192) & 0.387 (0.109) & 1.089 (0.302) & 1.197 (0.364) & \textbf{0.649 (0.053)} \\
    & 1 & 1.226 (0.158) & 1.005 (0.039) & 1.328 (0.231) & 1.087 (0.083) & \textbf{1.025 (0.033)} \\
    \hline
    & 0.2 & 0.764 (0.418) & 0.324 (0.204) & 0.654 (0.341) & 0.273 (0.260) & \textbf{0.174 (0.123)} \\
    0.002 & 0.5 & 1.133 (0.440) & 0.594 (0.452) & 1.222 (0.450) & 1.039 (0.440) & \textbf{0.782 (0.192)} \\
    & 1 & 1.555 (0.199) & 1.087 (0.101) & 1.449 (0.255) & 1.443 (0.294) & \textbf{1.142 (0.073)} \\
    \midrule\midrule
    \end{tabular}
    \label{table_5}
\end{table}
Table \ref{table_5} provides details. We try different sparsity values, $\pi=0.002$, $0.005$, $0.01$, and different signal strengths. We conduct the penalized regression method $10$ times for each sparsity and signal strength condition. Table \ref{table_5} summarizes mean values and standard deviations of ratios $\hat{\pi}/\pi$. The table also includes a column titled \textit{combination}. For each simulation, the maximum and minimum estimated proportion is dropped from the four penalized regression methods, and the mean is calculated from the remaining two estimates. Our \textit{combination} method provides a robust estimate of signal proportion and its standard deviation is always the lowest among all five methods across all conditions. While we generally use variance and bias to evaluate the estimator of a parameter, here, given the low standard deviation and its closeness to $1$, we use the \textit{combination} method estimator.

\subsection{Variable Selection Performance of Higher Criticism Tuned Regression}

Finally, to summarize the simulation studies, we combine all the steps in our proposed algorithm. We conduct variable selection performance between four current penalized regression methods and their corresponding higher criticism tuned regression. Table \ref{table_6} shows the results. Since the variance of the error term in our simulation model is $1$, $\beta_i=0.2$ or $0.5$ is relatively weak. This setting is for weak and sparse signals. In this scenario, it is challenging to find the true variables of a model. In Table \ref{table_6}, it is clear that for the original tuning region, $[0, \lambda_0]$, all four penalized regression methods select more variables than the true model. Especially for adaptive Lasso, the numbers of false positive variable selection are large for all four settings. As a comparison, when we update the tuning region as a higher criticism tuned region, $\{ \lambda:\floor*{\hat{\pi}p} \leq || \hat{\mathbf{\beta}}(\lambda) ||_0 \leq \ceil*{\hat{\pi}p} \}$, all four methods make tremendously fewer false positive variable selections. Some of the methods have similar performance because the most significant variables in a model are highly likely to be the same as those in another model. It is clear that after higher criticism tuning, the odds ratios are increased compared with original values. This means that both original and tuned methods work, while the tuned method provides more significant classifications. On the other hand, the $\text{F}_1$ score depends a lot on true positive choices, so in the weak signal environment, where $\beta = 0.2$, tuned methods returns lower scores. However, in the moderate signal environment, where $\beta = 0.5$, tuned methods show better classification performance again.

\begin{table}
    \caption[Method comparison between different penalized regression methods and higher criticism tuned regression]{Method comparison between different penalized regression methods and higher criticism tuned regression methods, with a sample size of $10$ across all methods, $p=1000$, $\sigma^2=1$.}
    \centering
    \begin{tabular}{cc|l|ccccccccc}
    \midrule\midrule
    $\pi$ & $\beta_{i}$ & Method & TP & FP & TN & FN & DOR & $\text{F}_1$ score \\
    \midrule\midrule
    \multirow{8}{*}{0.002} & \multirow{8}{*}{0.2} & Lasso & 0.6 & 13.5 & 984.5 & 1.4 & 31.254 & 0.075 \\
    & & \textbf{HC-Lasso} & 0.1 & 0.2 & 997.8 & 1.9 & 262.579 & 0.087 \\
    & & Adaptive Lasso & 1.1 & 74.9 & 923.1 & 0.9 & 15.063 & 0.028 \\
    & & \textbf{HC-Adaptive Lasso} & 0.1 & 0.2 & 997.8 & 1.9 &  262.579 & 0.087 \\
    & & SCAD $(\gamma = 3.7)$ & 0.4 & 4.6 & 993.4 & 1.6 &  53.989 & 0.114 \\
    & & \textbf{HC-SCAD} $(\gamma = 3.7)$ & 0.1 &  0.1 & 997.9 & 1.9 &  525.211 & 0.091 \\
    & & MCP $(\gamma = 3)$ & 0.2 & 1.1 & 996.9 & 1.8 &  100.697 & 0.121 \\
    & & \textbf{HC-MCP} $(\gamma = 3)$ & 0.1 & 0.1 & 997.9 & 1.9 &  525.211 & 0.091 \\
    \hline
    \multirow{8}{*}{0.002} & \multirow{8}{*}{0.5} & Lasso & 2.0 & 19.4 & 978.6 & 0.0  & 246.005 & 0.171 \\
    & & \textbf{HC-Lasso} & 1.2 & 0.0 & 998.0 & 0.8 & 2611.462 & 0.750 \\
    & & Adaptive Lasso & 2.0 & 47.7 & 950.3 & 0.0 & 98.631 & 0.077 \\
    & & \textbf{HC-Adaptive Lasso} & 1.1 & 0.0 & 998.0 & 0.9 & 2282.286 & 0.710 \\
    & & SCAD $(\gamma = 3.7)$ & 1.9 & 13.0 & 985.0 & 0.1 & 1439.615 & 0.225 \\
    & & \textbf{HC-SCAD} $(\gamma = 3.7)$ & 1.2 & 0.0 & 998.0 & 0.8 & 2611.462 & 0.750 \\
    & & MCP $(\gamma = 3)$ & 1.6 & 2.5 & 995.5 & 0.4 & 1592.800 & 0.525 \\
    & & \textbf{HC-MCP} $(\gamma = 3)$ & 0.9 & 0.1 & 997.9 & 1.1 & 8164.636 & 0.600 \\
    \midrule\midrule
    \multirow{8}{*}{0.005} & \multirow{8}{*}{0.2} & Lasso & 2.9 & 14.6 & 980.4 & 2.1 & 92.732 & 0.258 \\
    & & \textbf{HC-Lasso} & 0.4 & 0.0 & 995.0 & 4.6 & 351.353 & 0.148 \\
    & & Adaptive Lasso & 3.6 & 65.0 & 930.0 & 1.4 & 36.791 & 0.098 \\
    & & \textbf{HC-Adaptive Lasso} & 0.7 & 0.0 & 995.0 & 4.3 & 497.750 & 0.246 \\
    & & SCAD $(\gamma = 3.7)$ & 2.6 & 16.9 & 978.1 & 2.4 & 62.699 & 0.212 \\
    & & \textbf{HC-SCAD} $(\gamma = 3.7)$ & 0.5 & 0.0 & 995.0 & 4.5 & 398.200 & 0.182 \\
    & & MCP $(\gamma = 3)$ & 1.8 & 3.5 & 991.5 & 3.2 & 159.348 & 0.350 \\
    & & \textbf{HC-MCP} $(\gamma = 3)$ & 0.4 & 0.0 & 995.0 & 4.6 &  351.353 & 0.148 \\
    \hline
    \multirow{8}{*}{0.005} & \multirow{8}{*}{0.5} & Lasso & 5.0 & 17.9 & 977.1 & 0.0 & 584.435 & 0.358 \\
    & & \textbf{HC-Lasso} & 2.4 & 0.0 & 995.0 & 2.6 &  1862.548 & 0.649 \\
    & & Adaptive Lasso & 5.0 & 36.3 & 958.7 & 0.0 & 286.717 & 0.216 \\
    & & \textbf{HC-Adaptive Lasso} & 2.4 & 0.0 & 995.0 & 2.6 & 1862.548 & 0.649 \\
    & & SCAD $(\gamma = 3.7)$ & 4 & 9.7 & 985.3 & 1.0 & 406.309 & 0.428 \\
    & & \textbf{HC-SCAD} $(\gamma = 3.7)$ & 2.6 & 0.0 & 995.0 & 2.4 & 2128.31 & 0.684 \\
    & & MCP $(\gamma = 3)$ & 3.5 & 7.2 & 987.8 & 1.5 & 320.120 & 0.446 \\
    & & \textbf{HC-MCP} $(\gamma = 3)$ & 2.3 & 0.0 & 995.0 & 2.7 & 1742.125 & 0.630 \\
    \midrule\midrule
    \end{tabular}
    \label{table_6}
\end{table}

\section{Real Data Applications}
\label{sec:app}

We demonstrate our approach on two high dimensional use cases in human genetics.  The first is single-nucleotide polymorphism (SNP) data from a genome-wide association study evaluating low density lipoprotein (LDL) levels in patients with Type 2 diabetes \citep{marvel2017common}. The second is differential methylation data from an epigenome wide study evaluating differential methylation related to smoking pack years \citep{sikdar2019comparison}. The two different datatypes were chosen because they demonstrate the method on both discrete and continuous predictors. While both studies have genome-wide data available, candidate regions of substantial dimensionality were chosen based on previously established biological associations to aid in interpreting the results of our new approach.

\subsection{Discrete Predictor: Genome-Wide Association Study}

The Action to Control Cardiovascular Risk in Diabetes (ACCORD) clinical trial investigated whether intensive therapy to target normal glycated hemoglobin levels would reduce cardiovascular events in patients with type 2 diabetes \citep{action2008effects}. Genome wide SNP data was collected on a large proportion of patients within the trial, and the data has been used mapping a number of clinically important phenotypes \citep{tang2019genetic, morieri2018genetic, rotroff2018genetic1, shah2018modulation, rotroff2018genetic2, shah2016genetic}. Baseline lipid levels were the first trait mapped in ACCORD because the genetic etiology of lipid levels is well studied in non-diabetic people, with numerous studies and meta-analyses on hundreds of thousands of individuals replication \citep{marvel2017common}. We use the data for low density lipoprotein (LDL) levels to demonstrate our approach. Data from chromosome 19 was chosen for the current analysis because there are confirmed SNPs that are consistently associated with LDL levels.

Details of the data are described in \cite{marvel2017common}. Prior to analysis, quality control procedures were performed. For SNP level quality control, only SNPs with high variant call rates ($95\%$) and minor allele frequency ($5\%$) are kept. SNPs are also required to follow the expected proportions of Hardy-Weinberg equilibrium. A z-test for Hardy-Weinberg equilibrium is conducted for each SNP. If $|z_{\text{HWE}}| < |F^{-1}(0.5*10^{-6})|$, the SNP will be filtered. Following \cite{waldmann2013evaluation}, we replace missing alleles with the average allele frequency. For sample-level quality control, coefficients of inbreeding are calculated to filter samples with common ancestors. After quality control and genotype imputation, association of common genetic variants is tested using penalized multiple linear regression models and our proposed higher criticism tuned penalized regression models. In GWAS, the choice of covariates can significantly affect results. We use the $26$ phenotype observations employed in \cite{marvel2017common} as covariates of multiple linear regression models, including the top three principal components that summarize population substructure \citep{price2006principal}. Complete-case analysis was used. Finally, we have $n=4540$ individuals, $p=5746$ SNPs and $29$ covariates. There is no penalty effect put on these $29$ covariates, which means they will always remain in the model.  All four approaches were run on the data, and complete results are shown in Table \ref{table_7}-\ref{table_8}.

In Table \ref{table_7}, we calculate he estimated proportion to be $\hat{\pi}=0.00536=0.536\%$. Thus the estimated number of significant SNPs is $30.807$. Applying our estimation, we narrow the region for tuning parameter $\lambda$ to $[\lambda_{\text{min}}, \lambda_{\text{max}}]$, as shown in Table \ref{table_7}. While each method found a slightly different list of SNPs, with highly consistent results presented in bold. Here we highlight that all approaches identify rs445925, which is within the Apolipoprotein C1 (APOC1) gene. This SNP has been reported to be associated with LDL in previous publications, but was not identified when we tested for associations using traditional regression implemented locus-by-locus and then multiple testing procedures \citep{shatwan2018association, tang2015exome, deshmukh2012genome, trompet2011replication, marvel2017common}. APOC1 encodes a member of the apolipoprotein C1 family. It has been reported that its encoded protein plays a central role in high density lipoprotein (HDL) and very low density lipoprotein (VLDL) metabolism \citep{jong1999role, o2015reference}. This protein has also been shown to inhibit cholesteryl ester transfer protein in plasma \citep{de2009apolipoprotein, o2015reference}. Our discovery demonstrates the potential of the method to identify genetic association while considering correlation among biomarkers while higher power as compared to more traditional approaches. All selected SNPs are listed in Table \ref{table_8}. The selection differences among four methods are from the differences of solution paths for each respective penalized regression method.

\begin{table}
    \caption[HCR for GWAS on LDL from ACCORD]{Estimated proportion $\hat{\pi}$ and new tuning region $\{\lambda_{\text{min}}, \lambda_{\text{max}}\}$ for GWAS on baseline LDL and Chr $19$ from ACCORD data, where $p=5746$ and the number of covariate $q=29$.}
    \centering
    \begin{tabular}{l|c}
    \midrule\midrule
    \bf{Estimated Proportion} & $n=4540$ \\
    \hline
    $\hat{\pi}$ & $0.00536$ \\
    $\{\floor{\hat{\pi}p}, \ceil{\hat{\pi}p}\}$ & $\{30,31\}$ \\
    \midrule\midrule
    \bf{Models} & $\{\lambda_{\text{min}}, \lambda_{\text{max}}\}$ \\
    \hline
    Lasso & $\{1.218, 1.276\}$ \\
    Adaptive Lasso & $\{22.931, 24.022\}$ \\
    SCAD & $\{1.264, 1.303\}$ \\
    MCP & $\{1.226, 1.264\}$ \\
    \midrule\midrule
    \end{tabular}
    \label{table_7}
\end{table}

\begin{table}
    \caption[Selected SNPs in $n=4540$]{Selected SNPs in different regression models, $n=4540$. Overlapped SNPs in four models are in \textbf{bold}.}
    \centering
    \begin{tabular}{l|c|l}
    \midrule\midrule
    Model & $\lambda$ & Selected SNPs \\
    \hline
    \multirow{6}{*}{Lasso} & \multirow{6}{*}{$1.222$} & rs475192, \textbf{rs62106026}, \textbf{rs72999871}, \textbf{rs77071715}, \textbf{rs7254996}, \textbf{rs34002820}, \\
    & & \textbf{rs7255589}, \textbf{rs2106917}, rs10414987, \textbf{rs75981480}, \textbf{rs1687983}, rs4806078, \\
    & & \textbf{rs73037271}, \textbf{rs17265865}, \textbf{rs892594}, rs6859, \textbf{rs445925}, \textbf{rs11671132}, \\
    & & \textbf{rs251683}, \textbf{rs77059600}, \textbf{rs3745495}, rs11084080, rs2288868, rs10420138, \\
    & & \textbf{rs62110397}, \textbf{rs200657736}, rs3745902, \textbf{rs28506185}, rs8102873, \textbf{rs1548476}, \\
    & & \textbf{rs7257872} \\
    \hline
    \multirow{6}{*}{Adaptive Lasso} & \multirow{6}{*}{$22.931$} & rs263057, \textbf{rs62106026}, rs60357057, \textbf{rs72999871}, \textbf{rs77071715}, \textbf{rs7254996}, \\
    & & \textbf{rs34002820}, \textbf{rs7255589}, \textbf{rs2106917}, rs4806230, \textbf{rs75981480}, rs11668916, \\
    & & \textbf{rs1687983}, \textbf{rs73037271}, \textbf{rs17265865}, \textbf{rs892594}, \textbf{rs445925}, rs10426962, \\
    & & \textbf{rs11671132}, \textbf{rs251683}, \textbf{rs77059600}, \textbf{rs3745495}, rs73067324, rs10405559, \\
    & & \textbf{rs62110397}, \textbf{rs200657736}, \textbf{rs28506185}, \textbf{rs1548476}, rs10413455, rs55655541, \\
    & & \textbf{rs7257872} \\
    \hline
    \multirow{6}{*}{SCAD} & \multirow{6}{*}{$1.264$} & rs475192, \textbf{rs62106026}, \textbf{rs72999871}, \textbf{rs77071715}, \textbf{rs7254996}, \textbf{rs34002820}, \\
    & & \textbf{rs7255589}, \textbf{rs2106917}, rs10414987, \textbf{rs75981480}, \textbf{rs1687983}, rs4806078, \\
    & & \textbf{rs73037271}, \textbf{rs17265865}, \textbf{rs892594}, rs6859, \textbf{rs445925}, \textbf{rs11671132}, \\
    & & \textbf{rs251683}, \textbf{rs77059600}, \textbf{rs3745495}, rs11084080, rs2288868, rs10420138, \\
    & & \textbf{rs62110397}, \textbf{rs200657736}, rs3745902, \textbf{rs28506185}, rs8102873, \textbf{rs1548476}, \\
    & & \textbf{rs7257872} \\
    \hline
    \multirow{6}{*}{MCP} & \multirow{6}{*}{$1.231$} & rs475192, \textbf{rs62106026}, rs79038264, \textbf{rs72999871}, \textbf{rs77071715}, \textbf{rs7254996}, \\
    & & \textbf{rs34002820}, \textbf{rs7255589}, \textbf{rs2106917}, rs10414987, \textbf{rs75981480}, \textbf{rs1687983}, \\
    & & rs4806078, \textbf{rs73037271}, \textbf{rs17265865}, \textbf{rs892594}, \textbf{rs445925}, \textbf{rs11671132}, \\
    & & \textbf{rs251683}, \textbf{rs77059600}, \textbf{rs3745495}, rs11084080, rs2288868, rs10420138, \\
    & & \textbf{rs62110397}, \textbf{rs200657736}, rs3745902, \textbf{rs28506185}, rs8102873, \textbf{rs1548476}, \\
    & & \textbf{rs7257872} \\
    \midrule\midrule
    \end{tabular}
    \label{table_8}
\end{table}

\subsection{Continuous Predictor: DNA Methylation Data} \label{DNA_Methy}

In contrast to the categorical nature of the SNP data, differential methylation at Cytosine-phosphate-Guanine (CpG) sites across the genome is measured as a continuous trait that ranges from 0 to 1. As an application on continuous predictors, we evaluated data from an epigenome-wide association study (EWAS) for smoking related DNA methylation data in the Agricultural Lung Health Study, a case-control study of adult asthma nested within an agricultural cohort \citep{house2017early}. After data cleaning, our analysis includes IlluminaEPIC data from 2286 individuals. We investigate associations between pack-years, (packs smoked per day) $\times$ (years as a smoker), and methylation $\beta$ values at CpG sites in Chromosome $5$. This region was selected because it included 153 CpG sites mapped to the Aryl hydrocarbon receptor repressor (\textit{AHRR}) and the PDZ and LIM domain protein $7$ (\textit{PDLIM7}) genes (R package \textit{IlluminaHumanMethylationEPICmanifest}) that have validated associations across multiple studies \citep{joehanes2016epigenetic}. Additionally, to demonstrate our method on a similar level of dimensionality as the SNP data, we included selected CpG sites on the chromosome so that the number of sites, $p$, is 5000. 

Prior to analysis, quality control procedures were performed. We limit extreme values by winsorizing top and bottom $5\%$ $\beta$ values at each CpG site. Missing DNA methylation $\beta$ values are imputed by using observed average values. After quality control and imputation, association between pack year and DNA methylation $\beta$ values is tested using penalized multiple linear regression models and our proposed higher criticism tuned penalized regression models. Similar to GWAS, the choice of covariates also plays an important role in EWAS. We use $20$ variables (see Table \ref{table_covariate} for details) as covariates in multiple linear regression models, such as age, gender, principal components, and cell-type proportion. In summary, we have $n=2286$ individuals, $p=5000$ $\beta$ values 
and $20$ covariates. There is no penalty effect put on these $20$ covariates, which means they will always remain in the model. We conduct multiple EWAS with the cleaned sample $n=2286$ to evaluate our proposed approach. In Table \ref{table_9}, we calculate the estimated proportion to be $\hat{\pi}=0.000807=0.0807\%$. Thus the estimated number of significant CpG sites is $4.036$. Applying our estimation, we narrow the region for tuning parameter $\lambda$ to $[\lambda_{\text{min}}, \lambda_{\text{max}}]$, as shown in Table \ref{table_9}. All detected CpG sites are listed in Table \ref{table_10}. Those CpG sites in at least three models are in bold. The \textbf{cg05575921} site in the aryl hydrocarbon receptor repressor (AHRR) gene, a CpG gene consistently differentially methylated in relation to smoking across many studies, along with other CpGs in this gene \citep{monick2012coordinated, joehanes2016epigenetic}. We detected another well established CpG associated with smoking: \textbf{cg23576855} (also from AHRR gene). The association of \textbf{cg23576855} and smoking has been previously reported by \cite{philibert2012demethylation} and in the large meta-analysis of \cite{joehanes2016epigenetic}. This demonstrates the ability of our method, which considers correlation, to jointly detect well established associations in quantitative data.

\begin{table}
    \caption[HCR for EWAS on DNA Methylation Data]{Estimated proportion $\hat{\pi}$ and new tuning region $\{\lambda_{\text{min}}, \lambda_{\text{max}}\}$ for EWAS from DNA Methylation Data on Chr $5$, where $p=5000$ and the number of covariate $q=20$.}
    \centering
    \begin{tabular}{l|c}
    \midrule\midrule
    \bf{Estimated Proportion} & $n=2286$ \\
    \hline
    $\hat{\pi}$ & $0.0807\%$ \\
    $\{\floor{\hat{\pi}p}, \ceil{\hat{\pi}p}\}$ & $\{4,5\}$ \\
    \midrule\midrule
    \bf{Models} & $\{\lambda_{\text{min}}, \lambda_{\text{max}}\}$ \\
    \hline
    Lasso & $\{1.133, 1.498\}$ \\
    Adaptive Lasso & $\{1651.226, 1988.908\}$ \\
    SCAD & $\{1.123, 1.307\}$ \\
    MCP & $\{1.158, 1.268\}$ \\
    \midrule\midrule
    \end{tabular}
    \label{table_9}
\end{table}

\begin{table}
    \caption[Selected SNPs in $n=2286$]{Selected CpG sites in different regression models, $n=2286$. Overlapped CpG sites in at least three models are in \textbf{bold}.}
    \centering
    \begin{tabular}{l|c|l}
    \midrule\midrule
    Model & $\lambda$ & Selected CpGs \\
    \hline
    \multirow{1}{*}{Lasso} & \multirow{1}{*}{$1.133$} & \textbf{cg05575921}, \textbf{cg08916839}, cg11554391, \textbf{cg23576855}, \textbf{cg13032951} \\
    \hline
    \multirow{1}{*}{Adaptive Lasso} & \multirow{1}{*}{$1651.226$} & \textbf{cg05575921}, cg23953133, cg24965308, cg06016466, cg18174928 \\
    \hline
    \multirow{1}{*}{SCAD} & \multirow{1}{*}{$1.123$} & \textbf{cg05575921}, \textbf{cg08916839}, \textbf{cg23576855}, cg20075683, \textbf{cg13039251} \\
    \hline
    \multirow{1}{*}{MCP} & \multirow{1}{*}{$1.158$} & \textbf{cg05575921}, \textbf{cg08916839}, \textbf{cg23576855}, cg20075683, \textbf{cg13039251} \\
    \midrule\midrule
    \end{tabular}
    \label{table_10}
\end{table}

\section{Conclusions}
\label{sec:con}

For high-dimensional data, variable selection is always interesting but challenging. Penalized regression methods are preferred in high-dimensional setting because (1) they are more efficient than stepwise information criteria methods; and (2) they can consider effects of all variables simultaneously. The challenge is to find an appropriate penalty. In this article, we propose a new searching scheme for the regularization parameter $\lambda$ in penalized regression. We develop an algorithm to calculate \textit{p}-values in high-dimensional data through data multi-splitting and \textit{p}-value combination. We demonstrate that our approach is applicable to multiple penalized regression approaches. These results also compare the performance of each of the different penalized regression methods on genetic applications.

The estimator constructed with \textit{p}-values are used as a key step in our algorithm. We demonstrate the performance of our algorithm using both simulation tests and real data applications. In simulation tests, we prove that our method is robust for both weak and sparse signals, especially compared to other approaches. Using real data, we demonstrate the utility of our method in a real life setting. We analyze Chromosome 19 in ACCORD data and detect rs44592 whose association with LDL has been reported in previous publications. In epigenome-wide association study (EWAS), we detect association between smoking and $5,000$ CpG sites in Chromosome 5 and report some sites that were reported in other publications. Finally, we provide an R package on CRAN to implement our proposed approaches, which is available at \url{https://cran.r-project.org/web/packages/HCTR/index.html}.

\clearpage
\section{Supplementary Information}

\subsection{Covariates in Regression Model for DNA Methylation Data Applications}
\label{sec:covariate}

\begin{table}[h]
    \caption[Covariates in regression model for DNA methylation Data applications]{A list of covariates in regression model for DNA methylation data applications in Section \ref{DNA_Methy}}
    \centering
    \begin{tabular}{l|l|l}
    \midrule\midrule
    \textbf{Covariate} & \textbf{Unit} & \textbf{Note} \\
    \hline
    Age & Years & Numerical \\
    \hline
    Gender & Male/Female & Categorical \\
    \hline
    Body mass index & $kg/m^2$ & Numerical \\
    \hline
    Asthma status & Case/Noncase & Categorical \\
    \hline
    10 ancestry principal components &  & Numerical \\
    \hline
    Estimated cell type proportions: & & \multirow{3}{*}{Numerical} \\
    CD8+ and CD4+ T cells, NK cells, & & \\
    B cells, monocytes, granulocytes & & \\
    \midrule\midrule
    \end{tabular}
    \label{table_covariate}
\end{table}
The cell type proportions were estimated using the method described by \cite{houseman2012dna} with the reference panel intriduced by \cite{reinius2012differential}.

\clearpage
\bibliographystyle{apalike}  
\bibliography{references}  

\begin{thebibliography}{}

\bibitem[Ayers and Cordell, 2010]{ayers2010snp}
Ayers, K.~L. and Cordell, H.~J. (2010).
\newblock Snp selection in genome-wide and candidate gene studies via penalized
  logistic regression.
\newblock {\em Genetic Epidemiology}, 34(8):879--891.

\bibitem[Barnett et~al., 2017]{barnett2017generalized}
Barnett, I., Mukherjee, R., and Lin, X. (2017).
\newblock The generalized higher criticism for testing snp-set effects in
  genetic association studies.
\newblock {\em Journal of the American Statistical Association},
  112(517):64--76.

\bibitem[Breheny and Breheny, 2019]{breheny2019package}
Breheny, P. and Breheny, M.~P. (2019).
\newblock Package ‘ncvreg’.
\newblock {\em R package version}.

\bibitem[B{\"u}hlmann and Van De~Geer, 2011]{buhlmann2011statistics}
B{\"u}hlmann, P. and Van De~Geer, S. (2011).
\newblock {\em Statistics for high-dimensional data: methods, theory and
  applications}.
\newblock Springer Science \& Business Media.

\bibitem[Chinchor, 1992]{chinchor1992muc}
Chinchor, N. (1992).
\newblock Muc-4 evaluation metrics.
\newblock In {\em Proceedings of the 4th conference on Message understanding},
  pages 22--29. Association for Computational Linguistics.

\bibitem[Cho et~al., 2010]{cho2010joint}
Cho, S., Kim, K., Kim, Y.~J., Lee, J.-K., Cho, Y.~S., Lee, J.-Y., Han, B.-G.,
  Kim, H., Ott, J., and Park, T. (2010).
\newblock Joint identification of multiple genetic variants via elastic-net
  variable selection in a genome-wide association analysis.
\newblock {\em Annals of Human Genetics}, 74(5):416--428.

\bibitem[Consortium et~al., 2003]{complex2003nature}
Consortium, C.~T. et~al. (2003).
\newblock The nature and identification of quantitative trait loci: a
  community's view.
\newblock {\em Nature Reviews Genetics}, 4(11):911.

\bibitem[de~Barros et~al., 2009]{de2009apolipoprotein}
de~Barros, J.-P.~P., Boualam, A., Gautier, T., Dumont, L., Verg{\`e}s, B.,
  Masson, D., and Lagrost, L. (2009).
\newblock Apolipoprotein ci is a physiological regulator of cholesteryl ester
  transfer protein activity in human plasma but not in rabbit plasma.
\newblock {\em Journal of Lipid Research}, 50(9):1842--1851.

\bibitem[Deshmukh et~al., 2012]{deshmukh2012genome}
Deshmukh, H.~A., Colhoun, H.~M., Johnson, T., McKeigue, P.~M., Betteridge,
  D.~J., Durrington, P.~N., Fuller, J.~H., Livingstone, S., Charlton-Menys, V.,
  Neil, A., et~al. (2012).
\newblock Genome-wide association study of genetic determinants of ldl-c
  response to atorvastatin therapy: importance of lp (a).
\newblock {\em Journal of Lipid Research}, 53(5):1000--1011.

\bibitem[Dezeure et~al., 2015]{dezeure2015high}
Dezeure, R., B{\"u}hlmann, P., Meier, L., and Meinshausen, N. (2015).
\newblock High-dimensional inference: Confidence intervals, p-values and
  r-software hdi.
\newblock {\em Statistical Science}, pages 533--558.

\bibitem[Donoho et~al., 2004]{donoho2004higher}
Donoho, D., Jin, J., et~al. (2004).
\newblock Higher criticism for detecting sparse heterogeneous mixtures.
\newblock {\em The Annals of Statistics}, 32(3):962--994.

\bibitem[Dron and Hegele, 2016]{dron2016genetics}
Dron, J.~S. and Hegele, R.~A. (2016).
\newblock Genetics of lipid and lipoprotein disorders and traits.
\newblock {\em Current Genetic Medicine Reports}, 4(3):130--141.

\bibitem[Fan and Li, 2001]{fan2001variable}
Fan, J. and Li, R. (2001).
\newblock Variable selection via nonconcave penalized likelihood and its oracle
  properties.
\newblock {\em Journal of the American Statistical Association},
  96(456):1348--1360.

\bibitem[Friedman et~al., 2009]{friedman2009glmnet}
Friedman, J., Hastie, T., and Tibshirani, R. (2009).
\newblock glmnet: Lasso and elastic-net regularized generalized linear models.
\newblock {\em R package version}, 1(4).

\bibitem[Friedman et~al., 2010]{friedman2010regularization}
Friedman, J., Hastie, T., and Tibshirani, R. (2010).
\newblock Regularization paths for generalized linear models via coordinate
  descent.
\newblock {\em Journal of Statistical Software}, 33(1):1.

\bibitem[Gerstein et~al., 2008]{action2008effects}
Gerstein, H., Miller, M., Byington, R., Goff, D.~J., Bigger, J., Buse, J.,
  Cushman, W., Genuth, S., Ismail-Beigi, F., Grimm, R.~J., Probstfield, J.,
  Simons-Morton, D., and Friedewald, W. (2008).
\newblock Effects of intensive glucose lowering in type 2 diabetes.
\newblock {\em New England journal of medicine}, 358(24):2545--2559.

\bibitem[Glas et~al., 2003]{glas2003diagnostic}
Glas, A.~S., Lijmer, J.~G., Prins, M.~H., Bonsel, G.~J., and Bossuyt, P.~M.
  (2003).
\newblock The diagnostic odds ratio: a single indicator of test performance.
\newblock {\em Journal of Clinical Epidemiology}, 56(11):1129--1135.

\bibitem[Hall et~al., 2010]{hall2010innovated}
Hall, P., Jin, J., et~al. (2010).
\newblock Innovated higher criticism for detecting sparse signals in correlated
  noise.
\newblock {\em The Annals of Statistics}, 38(3):1686--1732.

\bibitem[House et~al., 2017]{house2017early}
House, J.~S., Wyss, A.~B., Hoppin, J.~A., Richards, M., Long, S., Umbach,
  D.~M., Henneberger, P.~K., Freeman, L. E.~B., Sandler, D.~P., O'Connell,
  E.~L., et~al. (2017).
\newblock Early-life farm exposures and adult asthma and atopy in the
  agricultural lung health study.
\newblock {\em Journal of Allergy and Clinical Immunology}, 140(1):249--256.

\bibitem[Houseman et~al., 2012]{houseman2012dna}
Houseman, E.~A., Accomando, W.~P., Koestler, D.~C., Christensen, B.~C., Marsit,
  C.~J., Nelson, H.~H., Wiencke, J.~K., and Kelsey, K.~T. (2012).
\newblock Dna methylation arrays as surrogate measures of cell mixture
  distribution.
\newblock {\em BMC bioinformatics}, 13(1):86.

\bibitem[Jeng et~al., 2019]{jeng2019efficient}
Jeng, X.~J., Zhang, T., and Tzeng, J.-Y. (2019).
\newblock Efficient signal inclusion with genomic applications.
\newblock {\em Journal of the American Statistical Association}, pages 1--23.

\bibitem[Joehanes et~al., 2016]{joehanes2016epigenetic}
Joehanes, R., Just, A.~C., Marioni, R.~E., Pilling, L.~C., Reynolds, L.~M.,
  Mandaviya, P.~R., Guan, W., Xu, T., Elks, C.~E., Aslibekyan, S., et~al.
  (2016).
\newblock Epigenetic signatures of cigarette smoking.
\newblock {\em Circulation: Cardiovascular Genetics}, 9(5):436--447.

\bibitem[Jong et~al., 1999]{jong1999role}
Jong, M.~C., Hofker, M.~H., and Havekes, L.~M. (1999).
\newblock Role of apocs in lipoprotein metabolism: functional differences
  between apoc1, apoc2, and apoc3.
\newblock {\em Arteriosclerosis, Thrombosis, and Vascular Biology},
  19(3):472--484.

\bibitem[Lockhart et~al., 2014]{lockhart2014significance}
Lockhart, R., Taylor, J., Tibshirani, R.~J., and Tibshirani, R. (2014).
\newblock A significance test for the lasso.
\newblock {\em Annals of Statistics}, 42(2):413.

\bibitem[Marvel et~al., 2017]{marvel2017common}
Marvel, S.~W., Rotroff, D.~M., Wagner, M.~J., Buse, J.~B., Havener, T.~M.,
  McLeod, H.~L., and Motsinger-Reif, A.~A. (2017).
\newblock Common and rare genetic markers of lipid variation in subjects with
  type 2 diabetes from the accord clinical trial.
\newblock {\em PeerJ}, 5:e3187.

\bibitem[Meinshausen et~al., 2009]{meinshausen2009p}
Meinshausen, N., Meier, L., and B{\"u}hlmann, P. (2009).
\newblock P-values for high-dimensional regression.
\newblock {\em Journal of the American Statistical Association},
  104(488):1671--1681.

\bibitem[Meinshausen et~al., 2006]{meinshausen2006estimating}
Meinshausen, N., Rice, J., et~al. (2006).
\newblock Estimating the proportion of false null hypotheses among a large
  number of independently tested hypotheses.
\newblock {\em The Annals of Statistics}, 34(1):373--393.

\bibitem[Monick et~al., 2012]{monick2012coordinated}
Monick, M.~M., Beach, S.~R., Plume, J., Sears, R., Gerrard, M., Brody, G.~H.,
  and Philibert, R.~A. (2012).
\newblock Coordinated changes in ahrr methylation in lymphoblasts and pulmonary
  macrophages from smokers.
\newblock {\em American Journal of Medical Genetics Part B: Neuropsychiatric
  Genetics}, 159(2):141--151.

\bibitem[Morieri et~al., 2018]{morieri2018genetic}
Morieri, M.~L., Gao, H., Pigeyre, M., Shah, H.~S., Sjaarda, J., Mendonca, C.,
  Hastings, T., Buranasupkajorn, P., Motsinger-Reif, A.~A., Rotroff, D.~M.,
  et~al. (2018).
\newblock Genetic tools for coronary risk assessment in type 2 diabetes: a
  cohort study from the accord clinical trial.
\newblock {\em Diabetes Care}, 41(11):2404--2413.

\bibitem[O'Leary et~al., 2015]{o2015reference}
O'Leary, N.~A., Wright, M.~W., Brister, J.~R., Ciufo, S., Haddad, D., McVeigh,
  R., Rajput, B., Robbertse, B., Smith-White, B., Ako-Adjei, D., et~al. (2015).
\newblock Reference sequence (refseq) database at ncbi: current status,
  taxonomic expansion, and functional annotation.
\newblock {\em Nucleic Acids Research}, 44(D1):D733--D745.

\bibitem[Olson et~al., 2017]{olson2017data}
Olson, R.~S., La~Cava, W., Mustahsan, Z., Varik, A., and Moore, J.~H. (2017).
\newblock Data-driven advice for applying machine learning to bioinformatics
  problems.
\newblock {\em arXiv preprint arXiv:1708.05070}.

\bibitem[Perneger, 1998]{perneger1998s}
Perneger, T.~V. (1998).
\newblock What's wrong with bonferroni adjustments.
\newblock {\em Bmj}, 316(7139):1236--1238.

\bibitem[Philibert et~al., 2012]{philibert2012demethylation}
Philibert, R.~A., Beach, S.~R., and Brody, G.~H. (2012).
\newblock Demethylation of the aryl hydrocarbon receptor repressor as a
  biomarker for nascent smokers.
\newblock {\em Epigenetics}, 7(11):1331--1338.

\bibitem[Price et~al., 2006]{price2006principal}
Price, A.~L., Patterson, N.~J., Plenge, R.~M., Weinblatt, M.~E., Shadick,
  N.~A., and Reich, D. (2006).
\newblock Principal components analysis corrects for stratification in
  genome-wide association studies.
\newblock {\em Nature Genetics}, 38(8):904.

\bibitem[Reinius et~al., 2012]{reinius2012differential}
Reinius, L.~E., Acevedo, N., Joerink, M., Pershagen, G., Dahl{\'e}n, S.-E.,
  Greco, D., S{\"o}derh{\"a}ll, C., Scheynius, A., and Kere, J. (2012).
\newblock Differential dna methylation in purified human blood cells:
  implications for cell lineage and studies on disease susceptibility.
\newblock {\em PloS one}, 7(7).

\bibitem[Rotroff et~al., 2018a]{rotroff2018genetic1}
Rotroff, D.~M., Pijut, S.~S., Marvel, S.~W., Jack, J.~R., Havener, T.~M.,
  Pujol, A., Schluter, A., Graf, G.~A., Ginsberg, H.~N., Shah, H.~S., et~al.
  (2018a).
\newblock Genetic variants in hsd17b3, smad3, and ipo11 impact circulating
  lipids in response to fenofibrate in individuals with type 2 diabetes.
\newblock {\em Clinical Pharmacology \& Therapeutics}, 103(4):712--721.

\bibitem[Rotroff et~al., 2018b]{rotroff2018genetic2}
Rotroff, D.~M., Yee, S.~W., Zhou, K., Marvel, S.~W., Shah, H.~S., Jack, J.~R.,
  Havener, T.~M., Hedderson, M.~M., Kubo, M., Herman, M.~A., et~al. (2018b).
\newblock Genetic variants in cpa6 and prpf31 are associated with variation in
  response to metformin in individuals with type 2 diabetes.
\newblock {\em Diabetes}, 67(7):1428--1440.

\bibitem[Shah et~al., 2016]{shah2016genetic}
Shah, H.~S., Gao, H., Morieri, M.~L., Skupien, J., Marvel, S., Par{\'e}, G.,
  Mannino, G.~C., Buranasupkajorn, P., Mendonca, C., Hastings, T., et~al.
  (2016).
\newblock Genetic predictors of cardiovascular mortality during intensive
  glycemic control in type 2 diabetes: findings from the accord clinical trial.
\newblock {\em Diabetes Care}, 39(11):1915--1924.

\bibitem[Shah et~al., 2018]{shah2018modulation}
Shah, H.~S., Morieri, M.~L., Marcovina, S.~M., Sigal, R.~J., Gerstein, H.~C.,
  Wagner, M.~J., Motsinger-Reif, A.~A., Buse, J.~B., Kraft, P., Mychaleckyj,
  J.~C., et~al. (2018).
\newblock Modulation of glp-1 levels by a genetic variant that regulates the
  cardiovascular effects of intensive glycemic control in accord.
\newblock {\em Diabetes Care}, 41(2):348--355.

\bibitem[Shatwan et~al., 2018]{shatwan2018association}
Shatwan, I.~M., Winther, K.~H., Ellahi, B., Elwood, P., Ben-Shlomo, Y., Givens,
  I., Rayman, M.~P., Lovegrove, J.~A., and Vimaleswaran, K.~S. (2018).
\newblock Association of apolipoprotein e gene polymorphisms with blood lipids
  and their interaction with dietary factors.
\newblock {\em Lipids in Health and Disease}, 17(1):98.

\bibitem[Sikdar et~al., 2019]{sikdar2019comparison}
Sikdar, S., Joehanes, R., Joubert, B.~R., Xu, C.-J., Vives-Usano, M., Rezwan,
  F.~I., Felix, J.~F., Ward, J.~M., Guan, W., Richmond, R.~C., et~al. (2019).
\newblock Comparison of smoking-related dna methylation between newborns from
  prenatal exposure and adults from personal smoking.
\newblock {\em Epigenomics}, 11(13):1487--1500.

\bibitem[Tang et~al., 2015]{tang2015exome}
Tang, C.~S., Zhang, H., Cheung, C.~Y., Xu, M., Ho, J.~C., Zhou, W., Cherny,
  S.~S., Zhang, Y., Holmen, O., Au, K.-W., et~al. (2015).
\newblock Exome-wide association analysis reveals novel coding sequence
  variants associated with lipid traits in chinese.
\newblock {\em Nature Communications}, 6:10206.

\bibitem[Tang et~al., 2019]{tang2019genetic}
Tang, Y., Lenzini, P.~A., Busui, R.~P., Ray, P.~R., Campbell, H., Perkins,
  B.~A., Callaghan, B., Wagner, M.~J., Motsinger-Reif, A.~A., Buse, J.~B.,
  et~al. (2019).
\newblock A genetic locus on chromosome 2q24 predicting peripheral neuropathy
  risk in type 2 diabetes: Results from the accord and bari 2d studies.
\newblock {\em Diabetes}, page db190109.

\bibitem[Tibshirani, 1996]{tibshirani1996regression}
Tibshirani, R. (1996).
\newblock Regression shrinkage and selection via the lasso.
\newblock {\em Journal of the Royal Statistical Society: Series B
  (Methodological)}, 58(1):267--288.

\bibitem[Tibshirani et~al., 2016]{tibshirani2016exact}
Tibshirani, R.~J., Taylor, J., Lockhart, R., and Tibshirani, R. (2016).
\newblock Exact post-selection inference for sequential regression procedures.
\newblock {\em Journal of the American Statistical Association},
  111(514):600--620.

\bibitem[Trompet et~al., 2011]{trompet2011replication}
Trompet, S., de~Craen, A.~J., Postmus, I., Ford, I., Sattar, N., Caslake, M.,
  Stott, D.~J., Buckley, B.~M., Sacks, F., Devlin, J.~J., et~al. (2011).
\newblock Replication of ldl gwas hits in prosper/phase as validation for
  future (pharmaco) genetic analyses.
\newblock {\em BMC Medical Genetics}, 12(1):131.

\bibitem[van~de Geer et~al., 2011]{van2011adaptive}
van~de Geer, S., B{\"u}hlmann, P., Zhou, S., et~al. (2011).
\newblock The adaptive and the thresholded lasso for potentially misspecified
  models (and a lower bound for the lasso).
\newblock {\em Electronic Journal of Statistics}, 5:688--749.

\bibitem[Waldmann et~al., 2013]{waldmann2013evaluation}
Waldmann, P., M{\'e}sz{\'a}ros, G., Gredler, B., Fuerst, C., and S{\"o}lkner,
  J. (2013).
\newblock Evaluation of the lasso and the elastic net in genome-wide
  association studies.
\newblock {\em Frontiers in Genetics}, 4:270.

\bibitem[Wasserman and Roeder, 2009]{wasserman2009high}
Wasserman, L. and Roeder, K. (2009).
\newblock High dimensional variable selection.
\newblock {\em Annals of Statistics}, 37(5A):2178.

\bibitem[Westfall et~al., 1993]{westfall1993resampling}
Westfall, P.~H., Young, S.~S., et~al. (1993).
\newblock {\em Resampling-based multiple testing: Examples and methods for
  p-value adjustment}, volume 279.
\newblock John Wiley \& Sons.

\bibitem[Wilson, 2019]{wilson2019harmonic}
Wilson, D.~J. (2019).
\newblock The harmonic mean p-value for combining dependent tests.
\newblock {\em Proceedings of the National Academy of Sciences},
  116(4):1195--1200.

\bibitem[Yu et~al., 2006]{yu2006unified}
Yu, J., Pressoir, G., Briggs, W.~H., Bi, I.~V., Yamasaki, M., Doebley, J.~F.,
  McMullen, M.~D., Gaut, B.~S., Nielsen, D.~M., Holland, J.~B., et~al. (2006).
\newblock A unified mixed-model method for association mapping that accounts
  for multiple levels of relatedness.
\newblock {\em Nature Genetics}, 38(2):203.

\bibitem[Zhang et~al., 2010a]{zhang2010nearly}
Zhang, C.-H. et~al. (2010a).
\newblock Nearly unbiased variable selection under minimax concave penalty.
\newblock {\em The Annals of Statistics}, 38(2):894--942.

\bibitem[Zhang et~al., 2010b]{zhang2010mixed}
Zhang, Z., Ersoz, E., Lai, C.-Q., Todhunter, R.~J., Tiwari, H.~K., Gore, M.~A.,
  Bradbury, P.~J., Yu, J., Arnett, D.~K., Ordovas, J.~M., et~al. (2010b).
\newblock Mixed linear model approach adapted for genome-wide association
  studies.
\newblock {\em Nature Genetics}, 42(4):355.

\bibitem[Zhao and Yu, 2006]{zhao2006model}
Zhao, P. and Yu, B. (2006).
\newblock On model selection consistency of lasso.
\newblock {\em Journal of Machine Learning Research}, 7(Nov):2541--2563.

\bibitem[Zhou and Stephens, 2012]{zhou2012genome}
Zhou, X. and Stephens, M. (2012).
\newblock Genome-wide efficient mixed-model analysis for association studies.
\newblock {\em Nature Genetics}, 44(7):821.

\bibitem[Zou, 2006]{zou2006adaptive}
Zou, H. (2006).
\newblock The adaptive lasso and its oracle properties.
\newblock {\em Journal of the American Statistical Association},
  101(476):1418--1429.

\end{thebibliography}


\end{document}